\documentclass[%
reprint,
superscriptaddress,
eprint,
twocolumn,
amsmath,
amssymb,
prx,
]{revtex4-2}
\usepackage[english]{babel}

\usepackage{graphicx}
\usepackage{dcolumn}
\usepackage{bm}
\usepackage{dsfont}
\usepackage[colorlinks]{hyperref}
\hypersetup{%
    plainpages=true,
    breaklinks=true,
    hypertexnames=false,
    pageanchor=true,
    colorlinks=true,
    linkcolor={blue},
    citecolor={blue},
    urlcolor={blue},
    anchorcolor={black}
}
\usepackage{hhline}
\usepackage{braket, siunitx}
\usepackage{ifthen}
\newboolean{showcomments}
\setboolean{showcomments}{true}

\usepackage{cleveref}
\usepackage{xcolor}
\usepackage{bbding} 
\usepackage{comment}
\usepackage{xspace}
\usepackage{hyperref}
\usepackage{soul}

\crefname{equation}{Eq.}{Eqs.}
\Crefname{equation}{Equation}{Equations}
\crefname{figure}{Fig.}{Figs.} 
\Crefname{figure}{Figure}{Figures}
\crefname{section}{Sect.}{Sects.}
\Crefname{section}{Section}{Sections}
\crefname{table}{Table}{Tables}
\crefname{appsec}{}{Appendices} 

\newcommand{\mynote}[3]{%
  \ifthenelse{\boolean{showcomments}}{%
   \fbox{\bfseries\sffamily\scriptsize#1}%
   {\small$\blacktriangleright$\textsf{\emph{\color{#3}{#2}}}$\blacktriangleleft$}}%
  {%
   \@bsphack
   \@esphack
  }%
}

\newcommand{\figlbl}{Fig.}
\newcommand{\Figlbl}{Figure}
\newcommand{\panel}[1]{({#1})}
\newcommand{\eqlbl}{Eq.}

\newcommand*{\figref}[2]{%
  \hyperref[{#1}]{%
    ~\ref*{#1}%
    \ifx\\#2\\%
    \else
      \panel{#2}%
    \fi
  }%
}

\newcommand{\EC}{E_\mathrm{C}}
\newcommand{\EJ}{E_\mathrm{J}}
\newcommand{\EL}{E_\mathrm{L}}
\newcommand{\Hop}{\hat H}
\newcommand{\Hopp}{{\hat H}_\mathrm{p}}
\newcommand{\phiop}{\hat\varphi}

\newcommand{\nop}{\hat n}
\newcommand{\mop}{\hat m}

\newcommand{\ngate}{n_{g}}

\newcommand{\phiext}{\varphi_\mathrm{ext}}

\newcommand{\psinkket}{|\psi_b^k\rangle}

\newcommand{\Phiext}{\Phi_\mathrm{ext}}
\newcommand{\TR}{T_{2\mathrm{R}}}
\newcommand{\TE}{T_{2\mathrm{E}}}
\newcommand{\TphiR}{T_{\varphi\mathrm{R}}}
\newcommand{\TphiE}{T_{\varphi\mathrm{E}}}
\newcommand{\TphiCQPS}{T_{\varphi \mathrm{R}}^\mathrm{CQPS}}
\newcommand{\GammaphiR}{\Gamma_{\varphi\mathrm{R}}}
\newcommand{\GammaphiE}{\Gamma_{\varphi\mathrm{E}}}
\newcommand{\GammaCQPS}{\Gamma_{\varphi\mathrm{R}}^\mathrm{CQPS}}
\newcommand{\zA}{z_{\mathrm{A}}}
\newcommand{\aA}{a_{\mathrm{A}}}
\newcommand{\cspec}{c_\mathrm{s}}
\newcommand{\Ecqps}{E_\mathrm{CQPS}}
\newcommand{\tauQP}{\tau_\mathrm{qp}}
\newcommand{\nQP}{N_\mathrm{qp}}
\newcommand{\jc}{J_\mathrm{c}}

\begin{document}
\preprint{APS/123-QED}
\widetext
\title{Dephasing in Fluxonium Qubits from Coherent Quantum Phase Slips}

\def\LLaffil{Lincoln Laboratory, Massachusetts Institute of Technology, Lexington, MA 02421, USA}
\def\RLEaffil{Research Laboratory of Electronics, Massachusetts Institute of Technology, Cambridge, MA 02139, USA}
\def\Physaffil{Department of Physics, Massachusetts Institute of Technology, Cambridge, MA 02139, USA}
\def\EECSaffil{Department of Electrical Engineering and Computer Science, Massachusetts Institute of Technology, Cambridge, MA 02139, USA}
\def\affilAQ{\textit{Atlantic Quantum, Cambridge, MA, USA}}
\def\affilGoogle{\textit{Google Quantum AI, Santa Barbara, CA, USA}}
\def\equalA{These authors contributed equally}

\author{Mallika T. Randeria}\altaffiliation[]{\equalA}\email{mallika.randeria@ll.mit.edu}\affiliation{\LLaffil}
\author{Thomas M. Hazard}\altaffiliation[]{\equalA}\email{thomas.hazard@ll.mit.edu}\affiliation{\LLaffil}
\author{Agustin~Di~Paolo}\altaffiliation[Present address: ]{\affilGoogle}\affiliation{\RLEaffil}
\author{Kate~Azar}\affiliation{\LLaffil}
\author{Max~Hays}\affiliation{\RLEaffil}
\author{Leon~Ding}\altaffiliation[Present address: ]{\affilAQ}\affiliation{\RLEaffil}\affiliation{\Physaffil}
\author{Junyoung~An}\affiliation{\RLEaffil}\affiliation{\EECSaffil}
\author{Michael~Gingras}\affiliation{\LLaffil}
\author{Bethany~M.~Niedzielski}\affiliation{\LLaffil}
\author{Hannah~Stickler}\affiliation{\LLaffil}
\author{Jeffrey~A.~Grover}\affiliation{\RLEaffil}
\author{Jonilyn~L.~Yoder}\affiliation{\LLaffil}
\author{Mollie~E.~Schwartz}\affiliation{\LLaffil}
\author{William~D.~Oliver}\affiliation{\RLEaffil}\affiliation{\Physaffil}\affiliation{\EECSaffil}
\author{Kyle~Serniak}\email{kyle.serniak@ll.mit.edu}\affiliation{\LLaffil}\affiliation{\RLEaffil}

\date{\today}
\begin{abstract}
Phase slips occur across all Josephson junctions (JJs) at a rate that increases with the impedance of the junction.
In superconducting qubits composed of JJ-array superinductors --- such as fluxonium --- phase slips in the array can lead to decoherence.
In particular, phase-slip processes at the individual array junctions can coherently interfere, each with an Aharonov--Casher phase that depends on the offset charges of the array islands.
These coherent quantum phase slips (CQPS) perturbatively modify the qubit frequency, and therefore charge noise on the array islands will lead to dephasing. 
By varying the impedance of the array junctions, we design a set of fluxonium qubits in which the expected phase-slip rate within the JJ-array changes by several orders of magnitude.
We characterize the coherence times of these qubits and demonstrate that the scaling of CQPS-induced dephasing rates agrees with our theoretical model.
Furthermore, we perform noise spectroscopy of two qubits in regimes dominated by either CQPS or flux noise.
We find the noise power spectrum associated with CQPS dephasing appears to be featureless at low frequencies and not $1/f$.
Numerical simulations indicate this behavior is consistent with charge noise generated by charge-parity fluctuations within the array.
Our findings broadly inform JJ-array-design tradeoffs, relevant for the numerous superconducting qubit designs employing JJ-array superinductors.  
\end{abstract}

\maketitle
\section{Introduction}
Superconducting qubits are a promising hardware platform for quantum computation \cite{kjaergaard_superconducting_2020, blais_circuit_2021}. 
Superconducting weak links, especially superconductor-insulator-superconductor Josephson junctions (JJs), are central to superconducting qubit design, often providing the nonlinearity necessary to define the qubit subspace within a spectrum containing many other non-computational states of the quantum circuit.
The strength of this nonlinearity is correlated with the magnitude of quantum fluctuations of the gauge-invariant phase difference of the superconducting order parameter across the JJ.
In the regime of large phase fluctuations, JJs can also undergo frequent ``phase slips"~\cite{likharev_theory_1985, zaikin_quantum_1997}---$2\pi$ phase discontinuities which necessarily preserve the single-valuedness of the macroscopic wavefunctions of its superconducting leads~\cite{tinkham_introduction_2015}. 
Here, the JJ is often thought of as a non-linear inductor. 
In a complementary role, arrays of Josephson junctions can be used to realize nearly linear inductors including superinductors with impedances approaching or exceeding the resistance quantum,~${R_\mathrm{Q}=h/(2e)^2\approx6.45~\mathrm{k}\Omega}$ at the gigahertz operating frequencies typical of superconducting qubits ~\cite{masluk_microwave_2012,bell_quantum_2012}. In this application, it is crucial that phase slips in the array are sufficiently rare.

Phase slips can be viewed as the dual process to Cooper-pair tunneling~\cite{mooij_superconducting_2006, kerman_fluxcharge_2013, koliofoti_compact_2023} and thereby underpin the dynamics of JJs in superconducting qubits and many other mesoscopic quantum devices.
Phase-slip processes have been shown to coherently interfere in arrays of Josephson junctions \cite{pop_measurement_2010, de_graaf_charge_2018, shaikhaidarov_quantized_2022}, a phenomenon referred to as Coherent Quantum Phase Slips (CQPS).
The coherent nature of this process implies very little intrinsic dissipation, and opens the possibility to realize a reproducible metrological link between current and frequency~\cite{likharev_theory_1985, crescini_evidence_2023} based on phase-slip elements~\cite{mooij_superconducting_2006}---the direct analog of voltage standards based on Josephson elements~\cite{hamilton_josephson_2000}.

The absence of dissipation is critical for the use of ``slippery" Josephson elements as building blocks of superconducting qubits.
Of particular note is the fluxonium qubit with its recently demonstrated state-of-the-art performance including coherence times surpassing one millisecond ~\cite{pop_coherent_2014, somoroff_millisecond_2023,ding_high-fidelity_2023}, as well as single- and two-qubit gate fidelities exceeding $99.99\%$ ~\cite{somoroff_millisecond_2023, ding_high-fidelity_2023} and 99.9\%~\cite{ding_high-fidelity_2023, zhang_tunable_2023}, respectively.
The fluxonium circuit is realized by shunting a single slippery JJ with a large superinductor~\cite{manucharyan_fluxonium_2009}, which effectively suppresses sensitivity to both low-frequency charge noise across the JJ, akin to the transmon~\cite{koch_charge-insensitive_2007} regime of the Cooper-pair-box circuit~\cite{bouchiat_quantum_1998}, and flux noise when compared to flux qubits with small inductive shunts~\cite{yan_flux_2016}.
This inductance can be realized in a number of modalities, including arrays of Josephson junctions~\cite{masluk_microwave_2012,bell_quantum_2012}, high-kinetic-inductance nanowires~\cite{hazard_nanowire_2019,grunhaupt_granular_2019,rieger_granular_2023}, or a geometric superinductor~\cite{peruzzo_surpassing_2020}. 
In a Josephson junction array (JJA), every junction provides a location for phase slips, and 
as was shown by Manucharyan~\emph{et al.}~\cite{manucharyan_evidence_2012}, a non-zero rate of CQPS in the JJA shunt inductor of a fluxonium qubit can reintroduce charge-noise-induced dephasing via fluctuating Aharonov--Casher-type~\cite{aharonov_topological_1984} interference.
More recently, theoretical efforts have explored mesoscopic models of this CQPS dephasing mechanism~\cite{mizel_right-sizing_2020,di_paolo_efficient_2021}, which motivate further experimental work to probe the charge dispersion in fluxonium. 

In this work, we expand upon the pioneering experiments of Ref.~\cite{manucharyan_evidence_2012} to quantify the contributions of CQPS dephasing in fluxonium qubits across a broad design space of junction-array superinductors.
We systematically tune the phase-slip rate by changing the impedance
of array junctions across six fluxonium qubits and show a scaling of CQPS dephasing rates over several orders of magnitude.
We characterize the fluctuations in qubit frequency to extract the noise power spectral density for two qubits in regimes limited either by flux noise or CQPS dephasing due to charge noise in the JJA.
Moreover, we perform numerical simulations that illustrate that charge-parity switching in the many islands of the JJA results in a Lorentzian-like noise power spectrum, consistent with our measurements for the CQPS-limited qubit.
Our experimental results corroborate theoretical models for CQPS-induced dephasing and reveal quantitative bounds for fluxonium design parameters in order to avoid this channel of decoherence.
The coupling of charge noise to the CQPS process has implications for superconducting quantum processors based on fluxonium and other noise-protected qubit circuits that rely on JJA superinductors~\cite{kou_fluxonium-based_2017,kalashnikov_bifluxon_2020,gyenis_experimental_2021}. 

\section{Theoretical Overview of CQPS Dephasing in Fluxonium}
\label{sec:Theory}
The CQPS dephasing mechanism is inherent to any superconducting qubit containing a closed loop in which phase slips can occur at multiple locations. Here, we quantitatively consider the specific case of a fluxonium qubit constructed with a JJA.
CQPS dephasing comprises several components: (i) quantum phase slips at the array junctions, or equivalently fluxon tunneling into and out of the superconducting loop of the qubit through the array, (ii) the Aharonov--Casher phase that the fluxon acquires from tunneling around offset charges on the array islands, (iii) the coherent addition of these geometric phase factors as multiple fluxon tunneling paths interfere, and (iv) the temporal fluctuations of these offset charges which broadens the qubit energy levels.
For completeness, we first discuss (A) phase slips in a single JJ, (B) the fluxonium circuit, (C) the CQPS process in a JJA, and (D) how it contributes to dephasing in fluxonium qubits. 

 \begin{figure}
    \centering
    \includegraphics[width=86mm]{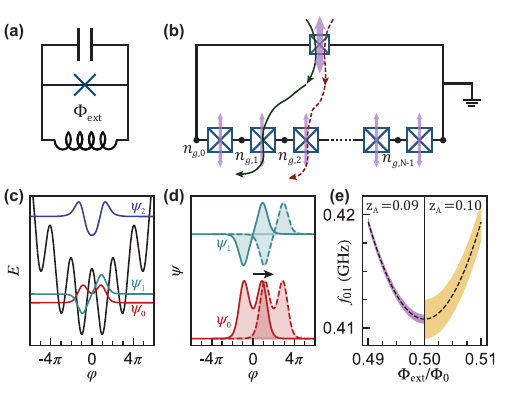}
    \caption{(a) Idealized fluxonium circuit, with a single JJ shunted by a linear inductor and capacitor.
    (b) An expanded fluxonium schematic where the inductor is implemented by a series array of JJs. 
    Phase slips are more likely to occur at the small junction than within the JJA (illustrated by the relative widths of the purple arrows).
    CQPS dephasing arises from series pairs of phase slip processes where one phase slip occurs at the small junction and the other occurs at one of the array junctions.
    Corresponding fluxon tunneling for two representative trajectories are shown by the green-solid and red-dashed lines.
    Such paths can interfere with different Aharonov--Casher phases depending on the total charge between the small junction and the particular array junction.
    The $j$-th superconducting island between JJs has an offset charge, $n_{g,j}\in [0,1)$ in units of Cooper-pair charge $2e$.
    (c)-(e) Computed fluxonium features using parameters $\EJ/h=3.2$~GHz, $\EC/h=1.4$ GHz and $\EL/h=0.25$ GHz.
    (c) Potential energy landscape (black curve) and corresponding fluxonium eigenstates at $\Phi_{\mathrm{ext}} = \Phi_0/2$.
    (d) Ground- and first-excited-state wavefunctions ($\psi_0$ and $\psi_1$, respectively) in the phase basis (offset vertically for visibility), and corresponding $2\pi$ displacements due to a phase slip in the array.
    (e) Calculated broadening of the fluxonium transition frequency $f_{01}$ near half flux due to CQPS.
    The shaded region indicates the standard deviation of the $f_{01}$ distribution around its unperturbed value (black dashed curve), shown for two different values of array-junction reduced impedance $\zA$.}
    \label{fig:1}
\end{figure}

\subsection{Phase slips in a Josephson junction}
When considering the dynamics of the phase degree of freedom across a Josephson junction, the periodicity of the superconducting order parameter allows for $2\pi$ phase slips to occur at the weak link. The magnitude of quantum fluctuations of the phase across a JJ is governed by its Josephson energy $\EJ$ and charging energy $\EC$. The phase-slip energy, $\epsilon_{\mathrm{ps}}$, for a JJ can be computed using the generalized WKB approximation \cite{matveev_persistent_2002, manucharyan_evidence_2012, di_paolo_efficient_2021}, and is given by
\begin{align}
    \epsilon_{\mathrm{ps}} &= 2\sqrt{\frac{2}{\pi}} \sqrt{8\EJ \EC}\sqrt[\leftroot{-1}\uproot{2}4]{\frac{8\EJ}{\EC}}\exp{\left( - \sqrt{\frac{8\EJ}{\EC}} \right)}\label{eq:epsilon0first} \\
    &= \frac{4 \sqrt{2}}{\pi}\hbar \omega_\mathrm{p} \sqrt{\frac{1}{z}} \exp{\left( - \frac{4}{\pi z} \right)}.
    \label{eq:epsilon0}
\end{align}
In the second line, we recast this equation in terms of the reduced junction impedance, a dimensionless quantity defined as $z \equiv~Z/R_\mathrm{Q}$, where $Z=~\sqrt{L_\mathrm{J}/C_\mathrm{J}}= (R_\mathrm{Q}/2\pi) \sqrt{8\EC / \EJ}$ is the impedance, $\hbar\omega_\mathrm{p} = \hbar/\sqrt{L_\mathrm{J}C_\mathrm{J}} = \sqrt{8\EJ \EC}$ is the plasma frequency, and  $L_\mathrm{J} = \Phi_0/(2\pi I_\mathrm{c})$ is the inductance for a JJ critical current $I_\mathrm{c}$. 

Crucially, this phase-slip energy is exponentially sensitive to the junction impedance. For small junction impedance, {\it i.e.} large $\EJ/\EC$, phase fluctuations are small. 
We also note the similarity between Eq.~\ref{eq:epsilon0first} and the charge dispersion of the Cooper-pair box/transmon circuit~\cite{koch_charge-insensitive_2007,koch_charging_2009}.

\subsection{Fluxonium}
The fluxonium qubit serves as the platform in which we explore these phase slips. The fluxonium circuit consists of a single high-impedance JJ (often called the ``small" JJ) shunted by a large inductor ~\cite{manucharyan_fluxonium_2009}, as shown schematically in \figlbl~\figref{fig:1}{a}. The circuit Hamiltonian is given by
\begin{equation}
   \hat{H} = 4 \EC \hat{n}^2 - \EJ \cos \hat\varphi + \frac{\EL}{2} \left(  \hat\varphi + 2\pi \frac{\Phiext}{\Phi_0} \right)^2,
   \label{eq:fluxoniumHamiltonian}
\end{equation}
where $\hat{\varphi}$ is an operator describing the phase difference across the small JJ, $\hat{n}$ is its conjugate variable for Cooper-pair number, $\Phiext$ is the external flux through the loop, and $\Phi_0 = h/2e$ is the superconducting flux quantum. The fluxonium~\cite{manucharyan_fluxonium_2009} resides in a regime of the generalized flux qubit circuit~\cite{yan_engineering_2020}
that supports low qubit operating frequencies, large anharmonicities~\cite{zhang_universal_2021, nguyen_high-coherence_2019}, protection against quasiparticle tunneling~\cite{pop_coherent_2014}, and non-trivial resonator-qubit interactions~\cite{smith_quantization_2016, zhu_circuit_2013,gusenkova_quantum_2021}.
Specifically, the Hamiltonian parameters must satisfy the criteria ${\EL \ll \EJ}$ and typically ${1 \lesssim \EJ/\EC \lesssim 10}$. 

Realizing small $\EL$ in physical devices can be challenging, and there are a few different approaches to construct the necessary large inductance~\cite{masluk_microwave_2012,bell_quantum_2012, hazard_nanowire_2019,grunhaupt_granular_2019,rieger_granular_2023, peruzzo_surpassing_2020}.
Here, we consider a superinductor that is implemented using an array of Josephson junctions, depicted schematically in \figlbl~\figref{fig:1}{b}. The relative ease of constructing JJAs and their process compatibility with standard qubit fabrication make them well suited for extensible qubit architectures. These arrays are typically formed with many ($N\gtrsim50$) JJs. For $N$ identical junctions, fluctuations of $\hat\varphi$ are uniformly distributed and therefore reduced by $1/N$ across each junction, enabling the linear approximation of the inductor with $\EL = E_\mathrm{JA}/N$, where $E_\mathrm{JA}$ is the Josephson energy for an individual array junction. 

Fluxonium qubits are often operated at ${\Phiext = \Phi_0/2}$, a ``sweet spot'' where the qubit is first-order insensitive to flux noise. At this flux bias point, the qubit frequency is dictated by the rate of phase slips across the small junction (typically in the tens or hundreds of MHz range).
We note that thus far, the degrees of freedom \emph{internal} to the JJA have been neglected. 

\subsection{CQPS in a Josephson-junction array}
The exponential sensitivity of the phase-slip rate on the ratio of a junction's Josephson and charging energies has implications for circuits with many JJs. In the case of fluxonium, the low impedance junctions in the JJA are constructed with large $E_\mathrm{JA}/E_\mathrm{CA}$, and therefore phase fluctuations across them are small compared to $2\pi$. Thus, phase slips at the array junctions are rare and can be treated as a perturbative correction to the fluxonium Hamiltonian~\cite{manucharyan_evidence_2012}
\begin{align}
    \Hop \approx 
    \sum_{\alpha} \epsilon_\alpha |\psi_\alpha \rangle \langle \psi_\alpha| + \underbrace{ \sum_{j\geq 1}^N \frac{\epsilon_{\mathrm{ps},j}}{2} e^{-i2\pi\eta_{g,j}} \mop^+  + \mathrm{h.c.}}_\text{perturbation},
    \label{eq: approx Hamiltonian}
\end{align}
where $\epsilon_\alpha$ and $|\psi_\alpha\rangle$ are the eigenvalues and eigenstates of the unperturbed Hamiltonian. Replacing the perfect inductor with a JJA makes it relevant to consider an offset charge across the small junction (\figlbl~\figref{fig:1}{b}), and include a charge offset in the first term of Eq.~\ref{eq:fluxoniumHamiltonian} (see ~\Cref{app:theory} for details).
Here, $\hat{m}^{+}$ is a phase-slip operator, and~$\epsilon_{\mathrm{ps},j}$ is the absolute value of the phase-slip amplitude corresponding to the $j$-th junction in the array.
This Hamiltonian is implicitly defined in terms of the winding-number operator associated with the phase, $\mop$, which equivalently represents the number of flux quanta stored in the loop.
More precisely, $\hat{\varphi} = -2\pi \hat{m} + \hat{\Omega}$, where~$\hat{\Omega}$ is the so-called interband potential~\cite{koch_charging_2009}. (The latter does not play a crucial role in the discussion that follows, but some of the details associated with it are discussed for completeness in~\Cref{app:theory} and \cite{di_paolo_manuscript_2024}.)
Equation~\ref{eq: approx Hamiltonian} incorporates two approximations within the perturbation term: it ignores plasmon excitations of the array junctions, and it assumes that no correlated phase-slip processes take place at a single array junction.
Both of these approximations are valid for typical junction-array parameters due to the large plasma frequency of the array junctions and their low impedance, which exponentially suppresses the probability of having more than one phase slip at a given junction. 

More importantly, Eq.~\ref{eq: approx Hamiltonian} has a simple interpretation: in the same way in which the small junction in fluxonium enables the coherent coupling of the persistent-current states~$|m\rangle$ ($\hat{m}$ eigenstates), the array junctions contribute to this process with a largely suppressed fluxon-tunneling amplitude, which is charge-offset dependent. Each superconducting island in the JJA can have an offset charge $n_{g,j}\in [0,1)$ in units of Cooper-pair charge, $2e$. Due to the Aharonov--Casher effect~\cite{aharonov_topological_1984, friedman_aharonov-casher-effect_2002}, the phase-slip amplitude at the $j$-th junction in the JJA picks up a geometric phase that depends on the total aggregated charge between the small junction and the $j$-th junction in the array where the phase slip occurred, $\eta_{g,j} = \sum_{k=0}^{j-1} n_{g,k}$.

The perturbative correction to the circuit eigenenergies are
\begin{align}
        \delta \epsilon_\alpha &= \sum_{j\geq 1}^N \frac{\epsilon_{\mathrm{ps},j}}{2} e^{-i2\pi\eta_{g,j}} \langle \psi_\alpha| \mop^+|\psi_\alpha\rangle + \mathrm{c.c.},
    \label{eq:perturbation theory correction}
\end{align}
to first-order in phase-slip amplitude.
It is instructive to separate out the eigenstate dependence of the equation above from the phase-slip components. 
We can define the total CQPS energy as 
\begin{equation}
    \Ecqps = \sum_{j=1}^N \epsilon_{\mathrm{ps},j}e^{-i 2\pi \eta_{g,j}},
    \label{eq:Escalc}
\end{equation}
which contains the phase slip energies for each junction in the array, $\epsilon_{\mathrm{ps},j}$, and the offset charge dependence of the Aharonov--Casher phase in the $\eta_{g,j}$ term. 

The matrix elements~$\langle \psi_\alpha| \mop^+|\psi_\alpha\rangle$ encompass the $2\pi$ displacement of the wavefunction generated by a phase slip through the array. 
While this term can be obtained in a numerically-exact fashion by representing the fluxonium Hamiltonian in the basis of eigenstates of~$\hat{m}$, a simpler expression follows from the approximation
\begin{equation}
    \langle \psi_\alpha| \mop^+|\psi_\alpha\rangle \approx  \langle \psi_\alpha | e^{-i2\pi\nop} |\psi_\alpha\rangle.
\end{equation}
The right-hand side of this equation links the matrix element of the phase-slip operator with that of a displacement operator, neglecting the contribution of the interband potential as an approximation. Crucially, however, $\langle \psi_\alpha | e^{-i2\pi\nop} |\psi_\alpha\rangle$ can be computed in any basis of choice for the fluxonium Hamiltonian, including, e.g. phase or Fock basis. 
For transitions between qubit states $\ket{\psi_\alpha}$ and $\ket{\psi_\beta}$, the difference between these matrix elements can be expressed in the phase basis as
\begin{align}
    \label{eq:structurefactor}
    \mathcal{F}_{\alpha \beta} & =  \langle \psi_\beta| \mop^+|\psi_\beta\rangle - \langle \psi_\alpha| \mop^+|\psi_\alpha\rangle\\
    & \approx \int_{-\infty}^{\infty} d\varphi \ \psi_{\beta}^*(\varphi) \psi_{\beta}(\varphi - 2\pi) \\
    & \ \ \ \ \  - \int_{-\infty}^{\infty} d\varphi \ \psi_{\alpha}^*(\varphi) \psi_{\alpha}(\varphi - 2\pi), \nonumber
\end{align}
where we refer to $\mathcal{F}_{\alpha \beta}$ as the ``structure factor" associated with this CQPS process, similar to the notation in \cite{manucharyan_evidence_2012}. Note $\psi_{\alpha}(\varphi)$ is the eigenstate of the unperturbed fluxonium Hamiltonian Eq.~\ref{eq:fluxoniumHamiltonian}.  \Figlbl\figref{fig:1}{c} illustrates the unperturbed fluxonium spectra at $\Phi_{\mathrm{ext}} = \Phi_0/2$, while \figlbl\figref{fig:1}{d} depicts the overlap of the shifted wavefunctions relevant for calculating the structure factor of the ground state and the first-excited state. This overlap function is determined by the circuit parameters and is maximal at $\Phi_{\mathrm{ext}} = \Phi_0/2$. 

Subtracting the ground- and first-excited state corrections in Eq.~\ref{eq:perturbation theory correction}, we arrive at the expression
\begin{equation}
    h\delta f_{01}  = \mathrm{Re}[\Ecqps \,\mathcal{F}_{01}],
    \label{eq: fab}
\end{equation}
which depends on the instantaneous value of $\Ecqps$, determined by the configuration of offset charges in the array. For $n_{g,0} = 0$, the structure factor is strictly real, and we recover ~$h\delta f_{01}  = \mathrm{Re}[\Ecqps ]\mathcal{F}_{01}$~\cite{manucharyan_evidence_2012}.

\subsection{Fluxonium dephasing from CQPS}
Since the correction to the unperturbed qubit frequency depends on the offset charges in the superinductor, charge noise gives rise to temporal fluctuations of the total CQPS amplitude $\Ecqps$. We now compute a bound on the CQPS-induced shift of the fluxonium transition frequency and
discuss how this effective linewidth translates to qubit dephasing. For a homogeneous JJA where $\epsilon_{\mathrm{ps},j} \equiv \epsilon_\mathrm{ps}$ for all $j \in [1, N]$, the total CQPS energy is restricted to a range $-N \epsilon_\mathrm{ps} \leq \mathrm{Re}[\Ecqps] \leq N\epsilon_\mathrm{ps}$.
However, the charges on each of the islands in the JJA are not typically controlled experimentally, so $\Ecqps$ will sample a smaller range of values generated by the time-dependent distribution of $\{ n_{g,j}\}$.
In order to mathematically convert from a noisy $\Ecqps$ to qubit frequency linewidth $\sigma_f$, we assume that $\Ecqps$ samples from a random distribution that varies slowly compared to the timescale of a measurement. This quasi-static assumption holds in the case of qubit frequency fluctuations that are slow compared to the Ramsey dephasing time $\TphiR$, such that the frequency is randomly sampling the Gaussian distribution during each instance of a Ramsey sequence.
We note that this assumption is not necessarily guaranteed, but is consistent with plausible models of local charge fluctuations, which we discuss later.

Given the $N \approx 100$ junctions in the array, we can use the central limit theorem to approximate the probability distribution of $\mathrm{Re}[\Ecqps]$ as a Gaussian with zero mean and standard deviation $\sqrt{N/2} \epsilon_\mathrm{ps}$. Consequently, the range of expected qubit frequencies is Gaussian distributed with a standard deviation $\sigma_f$ around its unperturbed value given by 
\begin{equation}
   h \left| \sigma_f \right| = \sqrt{\frac{N}{2}}\epsilon_\mathrm{ps} \left| \mathcal{F}_{01} \right|.
\end{equation}
To compute the corresponding dephasing rate, under the assumptions of quasi-static noise~\cite{ithier_decoherence_2005} and Gaussian-distributed qubit-frequency fluctuations, the Ramsey decay envelope will be given by the characteristic function of the distribution, resulting in a Gaussian decay with a dephasing rate   
\begin{equation}
    \GammaCQPS = \sqrt{2} \pi \left| \sigma_f \right| = \pi \sqrt{N} \epsilon_{\mathrm{ps}}\left| \mathcal{F}_{01} \right|.
    \label{eq:GammaCQPS}
\end{equation} 

\Figlbl\figref{fig:1}{e} illustrates the qubit frequency distribution for two different values of the array junction impedance $\zA = (1/2\pi) \sqrt{8 E_\mathrm{CA}/E_\mathrm{JA}}$, where $E_\mathrm{JA}$ and $E_\mathrm{CA}$ are the Josephson energy and single electron charging energy for an individual array junction. The shaded regions in \figlbl\figref{fig:1}{e} correspond to $\sigma_f$. In this plot, the Hamiltonian parameters $\EJ$, $\EC$, and $\EL$ are fixed and similar to those in our experimental devices.
Qualitatively, the rapidly increasing $\sigma_f$ for larger $\zA$ is due to the exponential sensitivity of the phase-slip amplitude $\epsilon_{\mathrm{ps}}$ to $\zA$ (\eqlbl~\ref{eq:epsilon0}). 
Interestingly, at the $\Phi_\mathrm{ext}=\Phi_0/2$ flux-noise sweet spot, the structure factor $\mathcal{F}_{01}$ and therefore $\GammaCQPS$ is \emph{maximal}, indicating that this bias point is a CQPS \emph{anti}-sweet spot.

\section{Experiment}
\label{sec:Experiment}
\subsection{Fluxonium device}
\begin{figure}
    \centering
    \includegraphics[width=86mm]{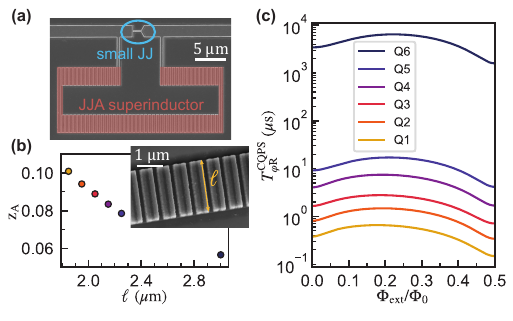}
    \caption{Fluxonium designs to probe CQPS dephasing.
    (a) Scanning electron micrograph of a fluxonium device with the small junction and its shunt JJA highlighted.
    (b) Inset: tilted scanning electron micrograph of the array junctions.
    The labeled dimension $\ell$ is varied in order to tune the fluxonium susceptibility to CQPS dephasing.
    Reduced impedance $\zA$ as a function of $\ell$ for an individual array junction, calculated using parameters for the six measured devices. The number of junctions in the array is adjusted along with $\ell$ in order to maintain the same nominal Hamiltonian parameters across all fluxoniums.
    (c) Theoretical CQPS dephasing time $\TphiCQPS$ as a function of $\Phi_\mathrm{ext}$, given by \eqlbl~\ref{eq:GammaCQPS}, plotted for the $\zA$ values shown in (b).}
    \label{fig:2}
\end{figure}

In this experiment, we focus on a single chip with six uncoupled planar fluxonium qubits. The qubits are designed to have nominally identical values for $\EJ$, $\EC$, and $\EL$, but the impedance of the array junctions is varied in order to engineer vastly different CQPS dephasing rates.
By keeping the Hamiltonian parameters approximately fixed, we probe CQPS dephasing absent other qubit-dependent variables.
An SEM image of one fluxonium is shown in \figlbl\figref{fig:2}{a}, where the small junction is shunted by a superinductance formed by an array of Josephson junctions. Each qubit is addressed by an independent microwave drive line and is capacitively coupled to an individual readout resonator, hanger-coupled to a shared microwave input-output line.
A global flux-biasing coil, mounted to the lid of the chip package, is used to tune $\Phi_{\mathrm{ext}}$ of the six fluxonium qubits simultaneously. As the mutual inductance between this coil and the individual qubits varies, the following experiments were serially performed on the different qubits. 

We explicitly change the phase-slip rate associated with the junctions in the array by varying their impedances $\zA = (1/2\pi) \sqrt{8 E_\mathrm{CA}/E_\mathrm{JA}}$, and compensate with the number of junctions in the JJA in order to keep ${\EL = E_\mathrm{JA}/N}$ fixed.
We achieve the requisite range of $\zA$ by changing the length $\ell$ of the array junctions (see dimension label in \figlbl\figref{fig:2}{b} inset), while the nominal width $w=200$~nm remains fixed in our Dolan-bridge-based~\cite{dolan_offset_1977} junction fabrication process.
For an individual array junction of area $\aA = \ell w$, its capacitance is given by $C = \cspec \aA$, where $\cspec$ is the specific capacitance (capacitance per unit area) associated with the JJ.
The Josephson energy also depends on junction geometry, with $\EJ = \Phi_0 I_\mathrm{c}/(2\pi)$ and $I_\mathrm{c} = J_\mathrm{c} \aA$, where $J_\mathrm{c}$ is the critical current density of the JJ.
For our device, we find $\cspec \approx 49$ fF/$\mu$m$^2$ (extracted from CQPS dephasing rates as described in the next subsection) and $J_\mathrm{c}\approx0.15\ \mu \mathrm{A}/\mu \mathrm{m}^2$ (extracted by matching Hamiltonian parameters to measured qubit spectra).
Consequently, the array junction impedance scales inversely with its area, as $\zA \propto 1/\aA$.
Thus, by varying the length of the junctions in the JJA, we control their impedance, as shown for our six qubits in  \figlbl\figref{fig:2}{b}. Table \ref{tab:Fluxonium Parameters} contains the full list of designed and measured parameters for each of the qubits on the chip.  

In \figlbl\figref{fig:2}{c} we plot the inverse of the expected CQPS dephasing rate, $\TphiCQPS$, calculated according to \eqlbl~\ref{eq:GammaCQPS}, for the six qubits using their $\zA$ shown in \figlbl\figref{fig:2}{b}.
The overall shape of $\TphiCQPS$ versus $\Phiext$ is indicative of the flux dependence of the structure factor $\mathcal{F}_{01}$ (\eqlbl~\ref{eq:structurefactor}). The magnitude of dephasing depends on the phase slip amplitude in the JJA, determined by $\zA$. We include a control qubit (Q6) that is designed to not be limited by CQPS dephasing, with an expected $\TphiCQPS$ that exceeds 1 ms over the entire flux range. 

\subsection{CQPS dephasing characterization}
\begin{figure*}
\centering
\includegraphics[width=170mm]{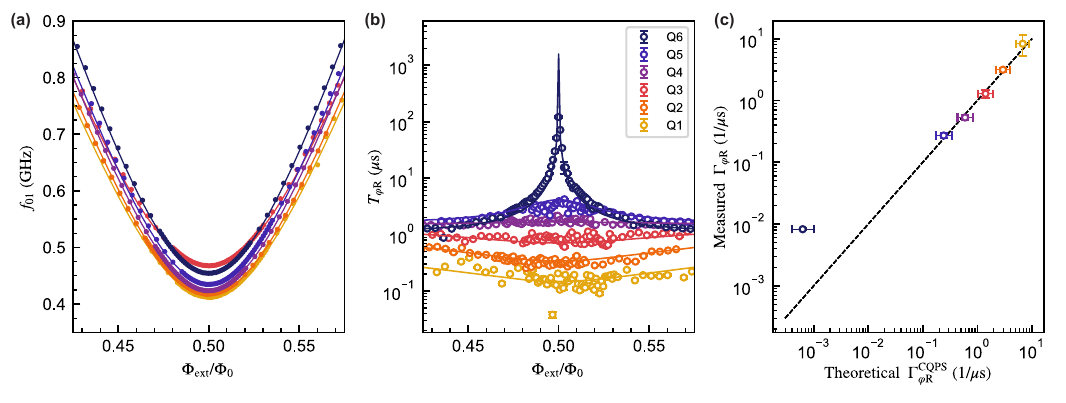}
\caption{(a)
Measured transition frequency $f_{01}$ (dots), overlaid with the calculated spectrum (lines), for the six fluxonium qubits studied in this experiment.
(b) Corresponding measured Ramsey dephasing times (open circles) and theoretical curves (solid lines), accounting for a combination of CQPS and first-order flux noise. The specific capacitance used to obtain the CQPS dephasing rates is fixed to be the same for all qubits at $\cspec = 49$ fF/$\mu$m$^2$. The flux noise amplitude is determined on a per-qubit basis from $\TE$ measurements and is not a free parameter in this model (Appendix \ref{app:fluxNoise}). (c) Measured Ramsey dephasing rates at $\Phiext = \Phi_0/2$ versus the theoretically expected dephasing rates from CQPS. The dashed line has unity slope; Q1 through Q5 match well to the expected CQPS scaling and Q6 was designed to {\it not} be limited by this mechanism. The $y$-axis data points and error bars for Q1 through Q5 are obtained from the average and standard deviation, respectively, of the seven points nearest $\Phiext = \Phi_0/2$. For Q6, we plot the the maximum measured $\TphiR$ value and corresponding fit uncertainty. The $x$-axis error bars correspond to $\GammaCQPS$ calculated for $\cspec = 49 \pm 2$ fF/$\mu$m$^2$.  }
\label{fig:3}
\end{figure*}

We first perform two-tone spectroscopy for each fluxonium over a large frequency and external flux range. We compare the measured transition frequencies to Hamiltonian simulations in order to extract the corresponding ~$\EJ$,~$\EC$, and~$\EL$ for each qubit (see Table \ref{tab:Fluxonium Parameters}). \Figlbl\figref{fig:3}{a} shows the measured spectroscopy around the $\Phi_{\mathrm{ext}} = \Phi_0/2$ operating point, where the transition frequency $f_{01}$ is approximately 450 MHz. All six qubits show similar transition spectra. This confirms that our device designs maintain similar Hamiltonian parameters, independent of expected CQPS dephasing rate. 

Next, we examine the fluxonium coherence times around half-flux as a probe of CQPS dephasing.
We characterize each qubit by measuring $T_1$, $\TR$, and $\TE$. In \figlbl\figref{fig:3}{b}, we plot $\TphiR$, the Ramsey pure-dephasing time, having accounted for the $T_1$-contribution to the decay function (see Appendix~\ref{app:exp details} for details).
Consistent with our assumed form of Gaussian noise in \eqlbl~\ref{eq:GammaCQPS}, the dephasing contribution to the Ramsey decay functions are fit with a Gaussian envelope.
The data show striking differences in $\TphiR$ across the set of qubits - both in terms of the shape of the curves as well as the coherence times around half-flux.
The distinctive peak in $\TphiR$ for the control qubit (Q6) is characteristic of the flux-noise sweet-spot, where the qubit is insensitive to first-order flux noise. In contrast, the other qubits (intended to be limited by CQPS dephasing) do not show that characteristic improvement of their coherence at half-flux, and in fact, exhibit a slight dip in $\TphiR$ due to the enhanced CQPS sensitivity at this bias point.

We compare this data to a model that includes both CQPS and flux-noise dephasing, with characteristic rates $\GammaCQPS$ and $\GammaphiR^{\Phi}$, respectively. We assume these two noise mechanisms are independent and Gaussian distributed; thus, the total dephasing rate is given by the quadrature sum of the two noise channels, $1/\TphiR = \GammaphiR = \sqrt{(\GammaCQPS)^2 + (\GammaphiR^{\Phi})^2}$.
The model incorporates the values of~$\EJ$,~$\EC$, and~$\EL$ extracted from spectroscopy and assumes the designed area for the junctions in the array.
For flux noise, we independently constrain $\GammaphiR$ by relating it to the qubit-specific $1/f$ flux noise amplitude obtained from fitting $\TE$ measurements, discussed in detail in Appendix~\ref{app:fluxNoise}.
Thus, our model for the total dephasing rate contains only one free parameter: the specific capacitance of the array JJs, which contributes to the $\GammaCQPS$ term.  Moreover, since $\cspec$ is determined solely by the fabrication process, we restrict it to be the same for all qubits, as it is not expected to vary significantly across the $5\times5$~mm chip.
We find that a value of $\cspec = 49$~fF/$\mu$m$^2$ shows good agreement with measured dephasing rates (solid lines in \figlbl\figref{fig:3}{b}).
Within these assumptions, we validate this minimal theoretical model for CQPS dephasing, with a {\it single} free parameter $\cspec$ across all six measured qubits.

\Figlbl\figref{fig:3}{c} shows measured values for $\GammaphiR$ at ${\Phiext=0.5~\Phi_0}$ compared to the theoretical prediction for dephasing arising solely from CQPS.
We see that for the five qubits designed to be limited by CQPS dephasing, our model accurately predicts $\GammaphiR$ across nearly two orders of magnitude.
The control qubit (Q6) is not limited by CQPS.
We suspect its coherence time at ${\Phiext=0.5~\Phi_0}$ also contains contributions from a combination of second-order flux noise \cite{ithier_decoherence_2005} and photon shot noise dephasing  ~\cite{sears_photon_2012}.
These results indicate that nearly identical Hamiltonian parameters (Eq.~\ref{eq:fluxoniumHamiltonian}) can have drastically different CQPS-limited dephasing times dictated by chosen parameters of the array junctions.
The implications this has for fluxonium designs, including JJA parameters, will be discussed in Sec.~\ref{sec:Outlook}.

\subsection{Noise spectroscopy}
To further validate our model and elucidate the origins of the charge noise leading to CQPS dephasing, it is insightful to compare the control qubit (Q6) -- with dephasing dominated by flux noise -- to qubit (Q1) -- with dephasing dominated most strongly by CQPS. In the following, we focus on measurements which probe different aspects of the dominant noise sources in these two fluxonium qubits. 

\begin{figure}
    \centering
    \includegraphics[width=86mm]{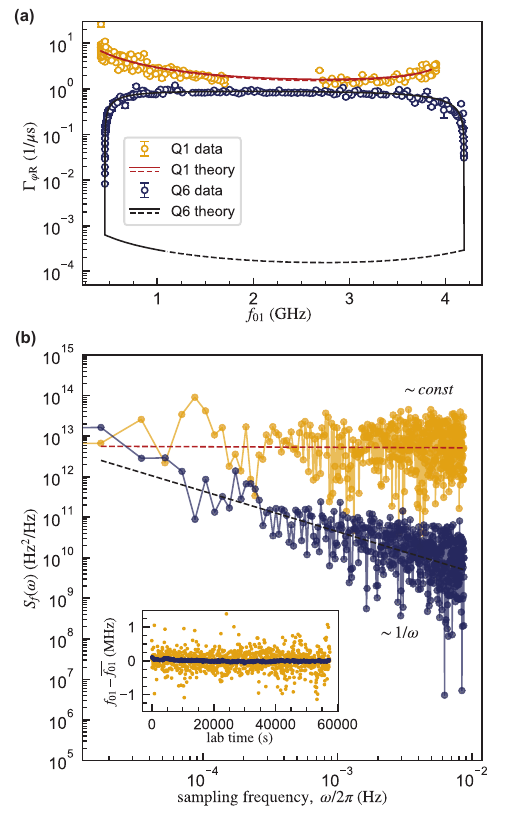}
    \caption{(a) Measured $\GammaphiR$ as a function of $f_{01}$ for two different qubits, Q1 (yellow dots) and Q6 (blue dots). Expected theoretical dephasing curves over the entire flux range, showing the contribution of CQPS dephasing only, $\GammaCQPS$  (dashed lines), and the combined effect of CQPS and flux noise, $\GammaphiR = \sqrt{(\GammaCQPS)^2 + (\GammaphiR^{\Phi})^2}$ (solid lines). Calculations use the respective JJA phase slip energies and scaled Ramsey flux noise amplitudes for Q1 (red) and Q6 (black). The total dephasing rates (measured data and solid lines)
    illustrates the difference in functional forms of dephasing dominated by CQPS (Q1) compared to flux noise (Q6). (b) Frequency power spectral density obtained from repeated measurements of $\TR$ for Q1 (yellow) and Q6 (blue), showing distinct noise behavior for each of the two qubits. Data from each qubit are fit to the equation $M/(\omega/2\pi)^{\mu}$, giving best fit values of $\mu = 0.00 \pm 0.06$ for Q1 (red dashed line), and $\mu = 1.00 \pm 0.06$ for Q6 (black dashed line). Extracted flux noise amplitude for Q6 is $A_{\Phi} \approx 5.9 \ \mu\Phi_0 / \sqrt{\mathrm{Hz}}$. Inset: qubit frequency fluctuations extracted from repeated Ramsey experiments, $f_{01}(t) - \overline{f_{01}}$, as a function of lab time, which are used to calculate the noise power spectrum. 
    }
    \label{fig:4}
\end{figure}

First, we examine the dependence of the dephasing rate over the entire range of external flux biases. \Figlbl\figref{fig:4}{a} highlights the strikingly different behavior for these two selected qubits. For the control qubit (Q6), $\GammaphiR$ evolves in a manner characteristic of flux noise, showing significant drops at the sweet spots (minimum and maximum fluxonium $f_{01}$), where the qubit becomes first-order insensitive to flux noise. In contrast, the qubit most susceptible to CQPS (Q1), exhibits the opposite behavior, with increased dephasing at the flux noise sweet-spots. This data corroborates that flux-noise sweet spots are anti-sweet-spots for CQPS-induced dephasing --- a feature that arises from the flux dependence of the structure factor $\mathcal{F}_{01}$ (\eqlbl~\ref{eq:structurefactor}).
In both cases, the data are well-captured over the entire flux range by the theoretical model which includes contributions from flux noise and CQPS dephasing, as discussed earlier.

Next we measure the noise power spectra of magnetic flux noise and charge noise leading to CQPS dephasing by characterizing fluctuations in qubit frequency across repeated measurements. 
To do this, we perform 1000 consecutive Ramsey measurements for qubits Q1 and Q6 at different $\Phiext$ bias points, where the dephasing is limited by these two noise sources.
We fit the oscillations in each Ramsey trace in order to obtain the qubit frequency; deviations in this frequency around its mean, $f_{01} (t) - \overline{f_{01}}$, are plotted as a function of laboratory time in the inset of \figlbl\figref{fig:4}{b}.
By examining correlations in these frequency fluctuations, we can obtain the noise power spectral density (PSD).  
This quantity, $S_f(\omega)$, shown in \figlbl\figref{fig:4}{b}, is calculated by taking the Fourier transform of the auto-correlation of $f_{01} (t) - \overline{f_{01}}$ and scaling by the sampling period. 

The extracted noise PSD for the two qubits have distinct functional forms.
To increase the measurement sensitivity to flux noise (ie. where $df_{01}/d\Phiext$ is large), we operate Q6 slightly away from half flux, at $\Phi_\mathrm{ext}/\Phi_0 \approx~0.42$. 
At this bias point, the measured $S_f(\omega)$ appears to be well-described by the $1/\omega$ 
functional form characteristic of flux noise~\cite{slichter_measurement-induced_2012,anton_pure_2012,braumuller_characterizing_2020,rower_evolution_2023}.
From the fit of the noise power spectrum, we obtain a flux noise amplitude of $A_{\Phi} = 5.9 \ \mu\Phi_0/ \sqrt{\mathrm{Hz}}$, similar to $A_{\Phi}$ extracted from  $\TE$ over the entire flux range for this qubit (Appendix~\ref{app:exp details}) and similar to values reported in other experiments \cite{slichter_measurement-induced_2012,bylander_noise_2011,nguyen_high-coherence_2019, braumuller_characterizing_2020, sun_characterization_2023-1}. 
This measurement protocol, similar to \cite{vepsalainen_improving_2022},
will yield a flux-noise amplitude slightly larger than that extracted from a single Ramsey decay, due to an extended low-frequency cutoff arising from the buffer time between subsequent Ramsey experiments.
\cite{ithier_decoherence_2005, smith_design_2019}.  

To probe the PSD of charge noise giving rise to CQPS dephasing, we repeat this measurement protocol for Q1 at $\Phiext = 0.5~\Phi_0$, where the qubit is maximally sensitive to charge fluctuations. In contrast to the flux-noise-limited Q6, for the CQPS-limited Q1 we extract a featureless noise PSD over our sampling frequency band, shown in \figlbl\figref{fig:4}{b}.
However, we find that $\TE$ values for this qubit show a significant increase compared to $\TR$ (see Appendix \ref{app:exp details}), indicating that the noise spectrum is not featureless at higher frequencies (up to $\sim1/\TE$).
Thus we posit the noise PSD has a distribution that is constant at low frequencies, but rolls off at higher frequencies.

\subsection{Noise modeling}
We propose a possible physical mechanism that could be responsible for a Lorentzian-like noise PSD due to charge noise in the array. Our noise spectroscopy results, discussed above, highlight that CQPS dephasing does not obey $1/f$ fluctuations. The charge on a superconducting island has two components, the continuous-valued offset charge and the discrete charge parity. Offset-charge drifts on small superconducting islands have been shown to have a $\sim1/f$ functional form ~\cite{ zimmerli_noise_1992, zorin_background_1996, kenyon_temperature_2000}.
Interestingly, it was recently shown that the charge noise sensed by larger superconducting islands can exhibit stronger time correlations, with PSDs scaling as $1/f^{\alpha}$ with $\alpha\approx 1.9$~\cite{christensen_anomalous_2019}.
As neither of these describe our data, we turn to considering the charge-parity switching within the array as the source of CQPS dephasing. 
The random telegraph noise associated with charge-parity jumps has been studied previously in other charge-sensitive circuits, including Cooper-pair transistors~\cite{naaman_poisson_2006} and offset-charge-sensitive transmons~\cite{riste_millisecond_2013,serniak_hot_2018, christensen_anomalous_2019, serniak_direct_2019, connolly_coexistence_2023}. This Poissonian process is known to give rise to a Lorentzian charge noise spectrum in those devices.

\begin{figure}[h]
    \centering
    \includegraphics[width=86mm]{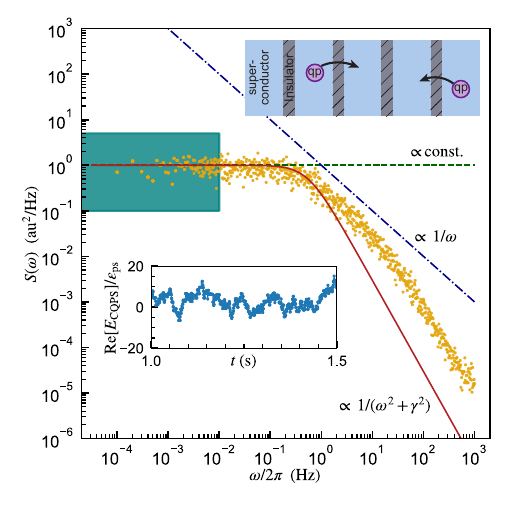}
    \caption{Simulations of charge-parity switching in a JJA. Top inset: schematic depicting quasiparticle tunneling, the mechanism which changes the charge parity of individual islands. Bottom inset: subset of a simulated time trace of $\mathrm{Re}[\Ecqps]$ normalized to $\epsilon_{\mathrm{PS}}$. Main panel: normalized noise power spectral density extracted from averaging ten different $10^4$ second-long simulated evolutions of $\mathrm{Re}[\Ecqps]$ (yellow points). For comparison, the functional forms for a Lorentzian lineshape (red line) characteristic of random telegraph noise, $1/\omega$ noise (blue line) and white noise (horizontal green line) are also shown. The green box highlights the sampling frequency range probed experimentally in \figlbl\figref{fig:4}{b}.}
    \label{fig:5}
\end{figure}

Here, we model charge noise within the JJA by simulating the time dynamics of discrete quasiparticle tunneling events between the superconducting islands of the array (upper inset of \figlbl\figref{fig:5}{}). We express the charge on the $j$-th island as $n_{\mathrm{g},j}(t)=n_{\mathrm{g},j}(0)+\delta n_{\mathrm{g},j}(t)+\frac{1}{2}\nQP{}_{,j}(t)$, in units of Cooper pair charge, $2e$. Here $n_{\mathrm{g},j}(0)$ is the initial offset charge, sampled randomly between $[0,1)$ and independently between islands; $\delta n_{g,j}$ is the slowly varying, continuous valued offset charge drift, and $\nQP{}_{,j}(t)$ is the discrete number of quasiparticles on the $j$-th island.
The characteristic timescale for quasiparticle tunneling, $\tauQP$, determines the probability for a charge-parity switch to occur in a time interval $\Delta t$ given by $1-\exp{(-\Delta t/\tauQP)}$. If the charge parity changes on the $j$-th island, a correlated charge-parity change is enforced on an adjacent island, in accordance with a quasiparticle tunneling to the left or right in the JJA (upper inset of Fig. \ref{fig:5}{}).

In the simulations shown in \figlbl\figref{fig:5}{}, we use the parameters for the JJA of Q1, and assume the same $\tauQP= 10$ ms between each of the islands. We note that the exact value of $\tauQP$ is not important for a qualitative match to the data in \figlbl\figref{fig:4}{b}, so long as $\sim1/\tauQP$ is larger than the maximum sampling frequency. We assign a fixed, random value of $n_{g,j}(0)$, and initialize the array with 10 quasiparticles distributed at random positions on the 
$N=85$ superconducting islands. 
After $\nQP{}_{,j}$ is calculated for every island at a particular time step, $\Ecqps$ is calculated in accordance with Eq.~\ref{eq:Escalc}, by coherently summing over all possible phase-slip trajectories and their enclosed charge.
The process is then repeated to obtain the time trace for $\nQP{}_{,j}(t)$ from which we calculate the simulated time dependence of $\Ecqps(t)$ (lower inset of Fig. \ref{fig:5}{}). In accordance with Eq.~\ref{eq: fab}, the real part of $\Ecqps$ is proportional to perturbative shift of the qubit frequency, and its time trace captures the temporal fluctuations of $\delta f_{01}$.
The power spectral density of the resulting $\Ecqps(t)$ is  obtained by taking the Fourier transform of the autocorrelation of $\Ecqps(t)$ and normalizing by the sampling period.
In \figlbl\figref{fig:5}{}, we plot the average of 10 simulations realized with different initial quasiparticle positions in the JJA, in order to capture the characteristic charge-parity switching behavior.
The simulations show a Lorentzian-like noise power spectrum that is constant at low frequencies and rolls off at a frequency below $1/\TE$, which is consistent with our repeated Ramsey and spin-echo measurements. 

Importantly, our results highlight that charge-parity switching, as opposed to $1/f$ charge noise, in the JJA is a plausible cause of CQPS dephasing. This work motivates further studies of charge noise in arrays of JJs, especially in the presence of various quasiparticle mitigation strategies \cite{riwar_normal-metal_2016, diamond_distinguishing_2022, marchegiani_quasiparticles_2022}. 

\begin{figure*}
    \centering
    \includegraphics{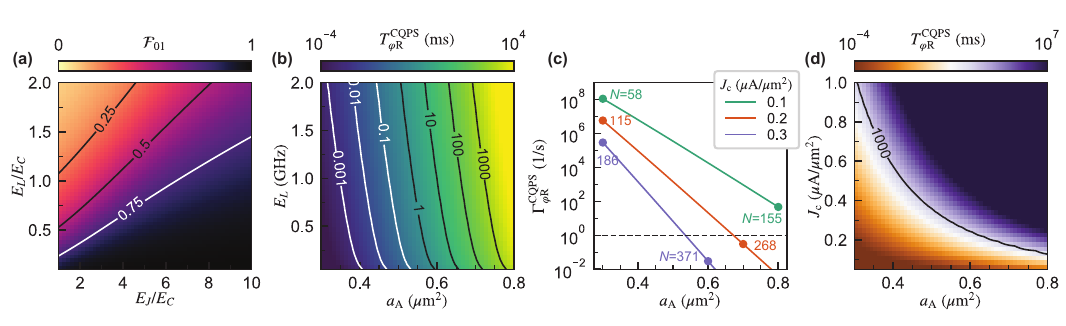}
    \caption{Tradeoffs within the design parameter space of fluxonium qubits with JJA superinductors. (a) The structure factor $\mathcal{F}_{01}$, corresponding to the overlap of the fluxonium wavefunctions displaced by $2\pi$ for different ratios of $E_L/E_C$ and $E_J/E_C$. (b) Ramsey-dephasing-time limit from CQPS for different values of $E_L$ and area of the array junctions $\aA$, allowing the number of junctions $N$ to vary in order to maintain the target $E_L$.  Here, $E_J/h=3.2$~GHz, $E_C/h=1.4$~GHz and $\jc=0.15 \ \mu\mathrm{A}/\mu\mathrm{m}^2$ are fixed. (c) CQPS-induced Ramsey dephasing rate for different $\jc$ values and varying $N$. Here, $E_J/h=3.2$~GHz, $E_C/h=1.4$~GHz and $E_L/h=0.25$~GHz are fixed. The dashed line indicates a $T_{\varphi} = 1$ s coherence limit. (d) Dephasing limit from CQPS over a range of typically accessible values for $\jc$ and $\aA$, with fixed target Hamiltonian parameters $E_J/h=3.2$~GHz, $E_C/h=1.4$~GHz, and $E_L/h=0.25$~GHz.}
    \label{fig:6}
\end{figure*}

\section{Outlook}
\label{sec:Outlook}
\subsection{Design considerations}
We now briefly consider some relevant design choices regarding arrays of JJs in order to mitigate CQPS dephasing in fluxonium qubits.
We summarize a possible workflow for fluxonium design accounting for CQPS dephasing due to charge noise in a JJA.
The first step is to choose a Hamiltonian of interest, which includes identifying the parameters $\EJ$, $\EC$, and $\EL$.
The target $\EJ$ and fabrication restrictions on junction dimensions and materials sets a range of reasonable $J_\mathrm{c}$ options.
The selected $J_\mathrm{c}$ and $\EL$ determine the size and number of array junctions since $\EL = E_\mathrm{JA}/N$.
Finally, within the parameter space of $\{\EL, J_\mathrm{c}, \aA, N \}$, one must ensure that CQPS dephasing is not a limiting factor for coherence of the qubit.

Given the wide range of parameter choices in a fluxonium design, we present several different parameterizations to examine tradeoffs within the design space.
The first consideration for CQPS dephasing is the structure factor, which is intrinsic to the Hamiltonian of interest and determined solely by the qubit wavefunction and its translation by $2\pi$ due to a phase slip, as given by Eq.~\ref{eq:structurefactor}.
\Figlbl\figref{fig:6}{a} shows the structure factor at $\Phi_{\mathrm{ext}} = \Phi_0/2$, and its dependence over a broad range of circuit parameters.
In cases where $\mathcal{F}_{01}$ is of order zero, the effect of CQPS dephasing will be negligible.
On the other hand, $\mathcal{F}_{01}$ is of order unity when $\EJ/\EC\gg1$ and $\EL/\EC\ll1$.
These limits include the heavy fluxonium regime~\cite{earnest_realization_2018,zhang_universal_2021}, a circuit under active research where careful consideration of CQPS dephasing would be valuable.

Next, we investigate the impact of JJA geometry on CQPS dephasing.
In \figlbl\figref{fig:6}{b}, we plot the theoretical estimate for $\TphiCQPS$ at $\Phi_{\mathrm{ext}} = \Phi_0/2$ as a function of $\EL$ and $\aA$, assuming $\EJ$, $\EC$ and $J_\mathrm{c}$ are fixed, with $N$ allowed to vary to achieve a particular $\EL$.
For a desired superinductance, one must fabricate large enough junctions for the JJA such that CQPS dephasing is not the limiting decoherence channel.
From the parameters chosen in \figlbl\figref{fig:6}{b}, array junctions with area greater than $\sim 0.7\ \mu \mathrm{m}^2$ are sufficient to maintain CQPS limited dephasing greater than 100 ms. 
\Figlbl\figref{fig:6}{c} illustrates the exponential scaling of $\GammaCQPS$ with $\aA$ for three different values of $J_\mathrm{c}$.
This calculation assumes $\EJ$, $\EC$ and $\EL$ are fixed, with the corresponding number of junctions $N$ in the array compensated accordingly to maintain the desired $\EL$.
The black dashed line in \figlbl\figref{fig:6}{c} corresponds to $\TphiCQPS = 1$ second, indicating that it is possible to choose realistic qubit parameters such that CQPS is not the limiting dephasing mechanism.

Finally, we examine the dependence of CQPS dephasing times on the two parameters commonly tuned by the design and fabrication process -  critical current density and array junction area.
The plot shown in \figlbl\figref{fig:6}{d} covers a relevant range of critical current densities often used in Al/AlOx/Al junction fabrication.
We find that CQPS dephasing can exceed 1 second over a large, physically realizable range of both $J_\mathrm{c}$ and $\aA$.  

Although the general trend from these different parameterizations point towards making arrays from many, larger, junctions rather than fewer, smaller junctions, it is worth cautioning against using this strategy arbitrarily.
Junction arrays have self-resonant modes, which can couple to the qubit and degrade coherence.
The density of these modes will increase with increasing $N$ due to increasing $\aA$~\cite{masluk_microwave_2012}.
A quantitative consideration of such array modes would require reliable estimates of stray capacitances to ground. Additional theoretical investigations have pointed to a potential reduction in $T_1$ for large $N$ arrays~\cite{mizel_right-sizing_2020}.
Moreover, as the size and number of array junctions is increased, the area of the loop formed by the array necessarily increases.
Previous experiments investigating the effects of flux noise on the geometry of superconducting loops embedded in qubits found an increase in the magnitude of flux noise, $A_\Phi$, with increasing loop perimeter~\cite{braumuller_characterizing_2020}.
For these reasons, it is desirable to choose Hamiltonian and junction parameters which ensure CQPS does not limit qubit coherence, while not pushing the array junction size and number too large.
By studying the CQPS dephasing mechanism in detail, we show that it is necessary to consider details of the JJA design beyond simply the Hamiltonian parameters, but it is possible to make design choices to ensure this is not the dominant noise channel.  

\subsection{Conclusions}
As the performance of superconducting qubits continues to improve, it is essential to understand all the mechanisms which can impact the performance of current and future devices.
Moreover, it can be valuable to revisit qubit design choices to ensure acceptable sensitivity to decoherence mechanisms. As an example, when the coherence times of transmon qubits improved beyond the first proof-of-principle demonstrations, relaxation via multi-mode Purcell effect~\cite{houck_controlling_2008} was found to limit coherence. This limitation has since been addressed sufficiently, though not removed entirely, by designing transmons with frequencies below that of the readout resonator and implementation of band-pass Purcell filters~\cite{reed_fast_2010, jeffrey_fast_2014}. In a similar vein, 
we systematically studied the effects of CQPS in fluxonium qubits specifically designed with varying sensitivity to this dephasing mechanism.
Our results verify the theoretical models necessary to design fluxonium qubits that sufficiently mitigate CQPS dephasing. 
Furthermore, we present measurements of the power spectral density for charge noise within a JJA inductor, and show qualitative agreement with a plausible model in which quasiparticle tunneling between the array junctions is the primary source of charge noise.
Although the presently reported CQPS dephasing can be improved by spin-echo sequences, mitigating this decoherence channel would avoid the additional overhead of algorithm-specific dynamical decoupling sequences necessary for high-fidelity qubit operations.
 
Our results confirm the predicted magnitude and scaling of CQPS dephasing over nearly two orders of magnitude in $\GammaphiR$ as the impedance $\zA$ of the array junctions is varied in fluxonium qubits. 
With this in mind, there appear to be no fundamental reasons why the coherence limit due to CQPS for typical fluxonium qubits cannot be extended beyond one second for practical experimental parameters, solely by reducing sensitivity to phase slips in the JJA via thoughtful design choices.
State-of-the-art coherence times for fluxonium are approximately 1 ms~\cite{somoroff_millisecond_2023}, indicating further improvements to flux-noise and dielectric-loss can be continued without hitting the CQPS coherence limit.
Separately, it is important to be mindful of this limitation of JJ arrays in the superinductance regime, as it pertains to other intrinsically-noise-protected superconducting qubits~\cite{gyenis_moving_2021}.

\section{Acknowledgements}
The authors thank Jeff Gertler, Ilan Rosen, Kunal Tiwari, and A. Jamie Kerman for insightful discussions. This research was funded in part under Air Force Contract No. FA8702-15-D-0001. L.D. gratefully acknowledges support from the IBM PhD Fellowship, J.A. gratefully acknowledges support from the Korea Foundation for Advances Studies. M.H. was supported by an appointment to the Intelligence Community Postdoctoral Research Fellowship Program at the Massachusetts Institute of Technology administered by Oak Ridge Institute for Science and Education (ORISE) through an interagency agreement between the U.S. Department of Energy and the Office of the Director of National Intelligence (ODNI). Any opinions, findings, conclusions or recommendations expressed in this material are those of the authors and do not necessarily reflect the views of the US Air Force or the US Government.

\appendix
\label{Appendix}
\section{Expanded experimental details}
\label{app:exp details}
\begin{figure*}[ht!]
    \centering
    \includegraphics{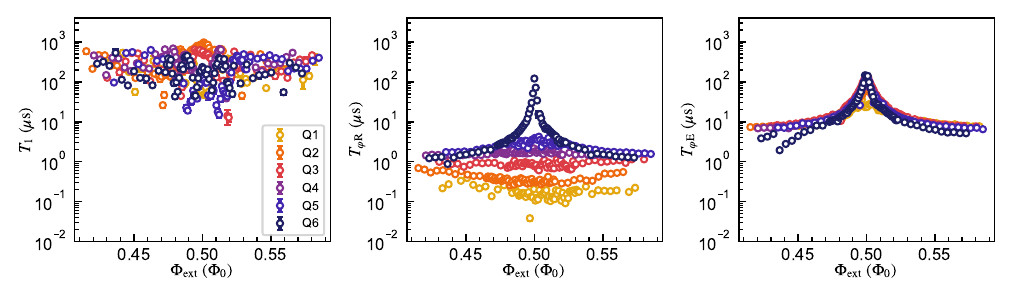}
    \caption{Measurements of $T_1$, $\TphiR$ and $\TphiE$ for all qubits near $\Phiext=0.5\Phi_0$.
    To increase readout visibility, the data were obtained with the pre- and post-measurement correlation scheme described in Appendix \ref{app:exp details}.
    The increase in $\TphiE$ compared to $\TphiR$ is most apparent for Q1-Q4, the qubits with the largest $\GammaCQPS$. }
    \label{fig:S1}
\end{figure*}

{\bf Qubit details:} Table \ref{tab:Fluxonium Parameters} contains additional device parameters extracted for each of the six qubits under test, such as the Hamiltonian parameters $\EJ$, $\EL$ and $\EC$, obtained from fitting two-tone spectroscopy data for the first three transitions of each qubit to Eq. \ref{eq:fluxoniumHamiltonian}.
Figure \ref{fig:S1} contains the $T_1$, $\TphiR$ and $\TphiE$ data taken near $\Phiext=\Phi_0/2$. All $T_2$ data for Ramsey and Echo dephasing are fit to a Gaussian decay using a functional form 
\begin{equation}
    f(t) = e^{-t/2T_1} e^{-(t/T_\varphi)^2} \cos (\omega t + \phi).
    \label{eq:gaussianDecay}
\end{equation}
In the fit, we use the measured $T_1$ value to account for the $T_1$ contribution. Fitting the pure dephasing to a Gaussian decay envelope is consistent within the bounds of $1/f$ flux noise, as well as for the case of quasi-static noise with Gaussian distribution of the qubit frequency fluctuations discussed in the main text. 

We emphasize that flux noise and CQPS are both expected to independently result in Gaussian dephasing envelope $\propto e^{-(t/T_\varphi)^2}$. Therefore, the net dephasing rate is given by the quadrature sum of dephasing from these two mechanisms, $(\Gamma_{\varphi}^{\mathrm{total}})^2 = (\GammaCQPS)^2 + (\GammaphiR^{\Phi})^2$.

We suspect the variation in $T_1$ values around half-flux is likely due to the presence of spurious two level systems (TLS) modes that vary from qubit to qubit, consistent with what is seen by many groups \cite{sun_characterization_2023-1}. We note that the shape of $\TphiE$ measurements versus flux indicate the Echo coherence times are limited by flux noise. Moreover, the enchancement in $\TphiE$ compared to $\TphiR$, are consistent with a noise power spectral density for CQPS dephasing that is constant at low frequencies but rolls off at a frequency below the echo filter function.

\renewcommand{\arraystretch}{1.5}
\begin{table*}[ht]
\centering
\begin{tabular}{c||c|c|c|c|c|c} 
 & Q1 & Q2 & Q3 & Q4 & Q5 & Q6 \\
 \hline
 \emph{Design parameters} & & & & & & \\
Number of array JJs: $N$ & 85 & 89 & 95 & 99 & 103 & 139\\
Length of array JJs: $\ell$ ($\mu$m) & 1.85 & 1.95 & 2.05 & 2.15 & 2.25 & 3.00\\
 \hline
\emph{Inferred from measurement} & & & & & & \\
$\EJ/h$ (GHz) & 3.22 & 3.165 & 3.1 & 3.315 & 3.13 & 3.2 \\ 
$\EC/h$ (GHz) & 1.41 & 1.38 & 1.45 & 1.43 & 1.37 & 1.39 \\ 
$\EL/h$ (GHz) & 0.25 & 0.26 & 0.26 & 0.27 & 0.28 & 0.3 \\ 
\hline
\emph{Measured values at half-flux} & & & & & & \\
$f_{01}$ (GHz) & 0.412 & 0.417 & 0.467 & 0.423 & 0.435 & 0.454 \\
$T_1$ ($\mu$s) & 56.5 $\pm$ 6.0 & 885.9 $\pm$ 79.8 & 579.1 $\pm$ 45.6 & 165.0 $\pm$ 41.4 & 126.4 $\pm$ 14.1 & 235.6 $\pm$ 19.6 \\
$\TphiR$ ($\mu$s) & 0.21 $\pm$ 0.01 & 0.31 $\pm$ 0.01 & 0.76 $\pm$ 0.01 & 2.16 $\pm$ 0.06 & 3.93 $\pm$ 0.06 & 121.1 $\pm$ 4.8\\
$\TphiE$ ($\mu$s) & 23.5 $\pm$ 0.5 & 62.0 $\pm$ 1.1 & 132.6 $\pm$ 3.4 & 85.6 $\pm$ 3.2 & 140.1 $\pm$ 5.9 & 142.0 $\pm$ 5.4\\
\hline
\emph{From fits} & & & & & & \\
$z_A$ for $\cspec$ = 49 fF/$\mu$m$^2$ & 0.101 & 0.094 & 0.089 & 0.083 & 0.079 & 0.057\\
$A_{\Phi}$ ($\mu\Phi_0 / \sqrt{\mathrm{Hz}}$) from $\TphiE$ & 3.35 & 3.30 & 3.50 & 3.45 & 3.30 & 4.5\\
\end{tabular}
\caption{Design parameters for the JJA, extracted Hamiltonian parameters, measured coherence times at $\Phiext=0.5\Phi_0$, as well as reduced array junction impedances and flux noise amplitudes for each of the six fluxonium qubits. 
}
\label{tab:Fluxonium Parameters}
\end{table*}

\begin{figure*}
    \centering
    \includegraphics{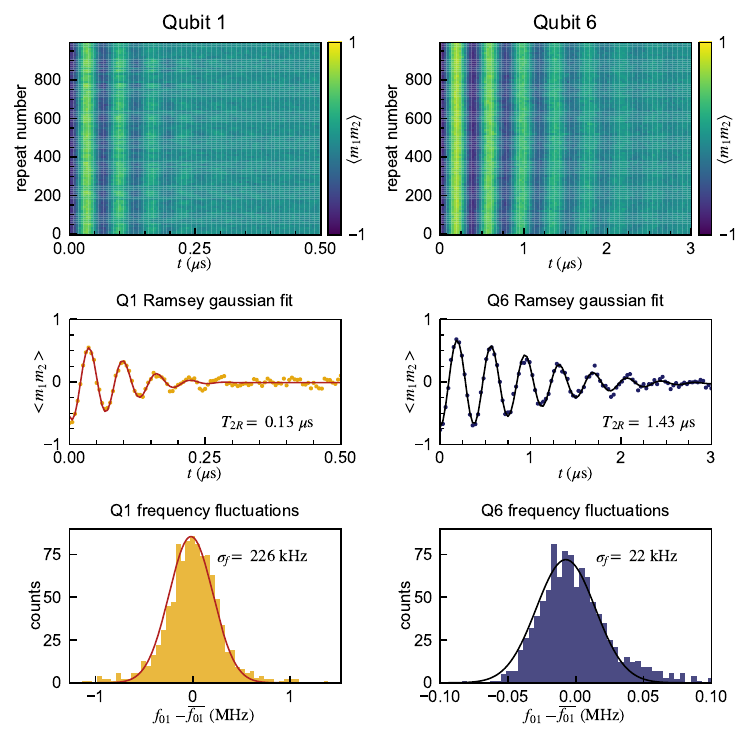}
    \caption{Top panels: Every tenth trace of 1000 repeated Ramsey measurements for Q1 (left) and Q6 (right). Middle panels: data from an individual Ramsey experiment for each qubit and corresponding fits with Gaussian decay envelopes. Lower panels: histograms of frequency fluctuations for each qubit and corresponding fit to a Gaussian distribution with zero mean. Q1 is measured at $\Phiext = \Phi_0/2$, Q6 is measured at $\Phiext = 0.42\Phi_0$ to enhance sensitivity to CQPS dephasing and flux noise dephasing, respectively.}
    \label{fig:S3}
\end{figure*}

{\bf Data acquisition:}  For qubits with long $T_1$ times or significant excited-state population at low qubit frequencies, adding an additional measurement pre-pulse before a pulse sequence and comparing the pre- and post-measurement results (denoted $m_1$ and $m_2$, respectively) can significantly increase readout signal visibility and decrease the time between successive pulse sequences, i.e., to avoid waiting several $T_1$ for the qubit to return to thermal equilibrium.
To that end, where $f_{01}< 1$ GHz, the data were acquired with said pre- and post-measurement sequence which results in near unity visibility, up to measurement error due to $T_1$ events during each of the two measurements.
The $I$ and $Q$ quadrature voltages from each measurement are binned into the either the $|0\rangle$ and $|1\rangle$ qubit states and then compared to each other to obtain the ensemble-averaged measurement correlation function, $\langle m_1m_2\rangle$. An example is shown for $\TR$ measurements in the lower panels of \Figlbl\figref{fig:S3}{}.
For $f_{01}> 1$ GHz, the equilibrium excited-state population is sufficiently small and the readout visibility is large enough such that a single post-sequence measurement results in an acceptable signal-to-noise ratio.

\section{Flux noise}
\label{app:fluxNoise}

\begin{figure*}
    \centering
    \includegraphics{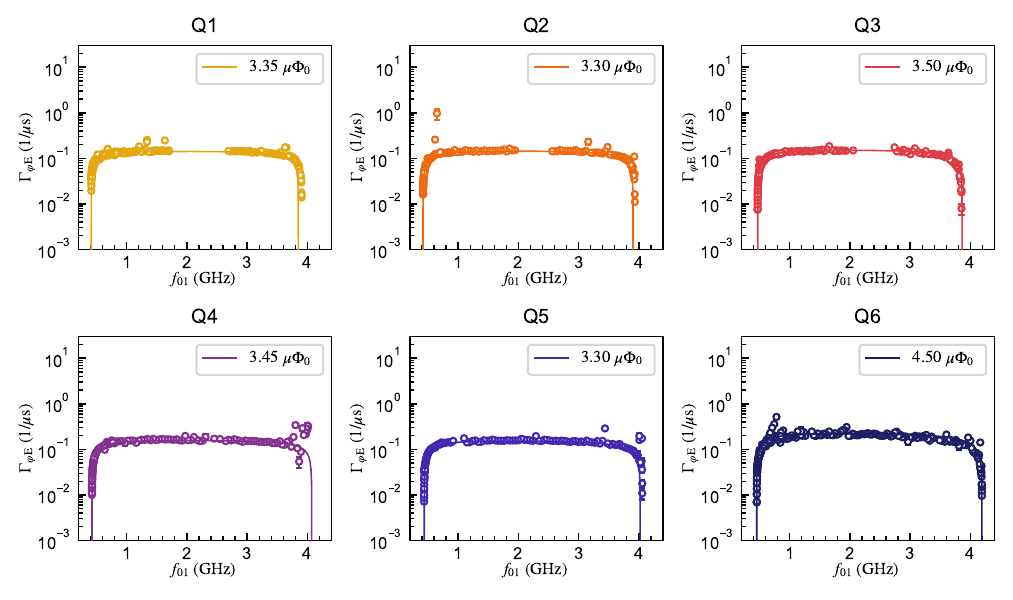}
    \caption{The measured values of $\GammaphiE$ for all six qubits over the entire range of external flux biases, plotted at their corresponding qubit frequencies $f_{01}$.  The solid line is a fit to a dephasing model that only includes first-order flux noise, with the corresponding value of $A_\Phi$ listed for each qubit in the plot legend.}
    \label{fig:S2}
\end{figure*}

We write the power spectral density for $1/f$ flux noise as
\begin{equation}
    S_{\Phi} (\omega) = A_{\Phi}^2 \left( \frac{2\pi \times 1 \mathrm{Hz}} {|\omega|} \right),
\end{equation}
where $A_{\Phi}$ is the flux noise amplitude. Away from the flux sweet spots, the qubit is sensitive to first-order perturbations of the flux bias ($\partial\omega_{01}/\partial\Phi_\mathrm{ext}\neq 0$). The decay envelope of the off-diagonal term of the qubit density matrix due to  $1/f$ noise is expected to be Gaussian with 
a dephasing rate given by
\begin{equation}
     \Gamma_{\varphi}^{\Phi}  = \left| \frac{\partial\omega_{01}}{\partial\Phiext} \right| A_{\Phi} \sqrt{\xi},
\end{equation}
where $\xi$ is a numerical factor characteristic of the filter function of a particular measurement sequence \cite{ithier_decoherence_2005,smith_design_2019}. The filter function governing the free induction decay in a Ramsey experiment diverges at low frequencies and therefore the Ramsey decay rate depends on the infrared cutoff $\omega_\mathrm{ir}$ (set by the total duration of the Ramsey experiment), with $\xi_{\mathrm{R}} \approx \mathrm{ln} (1/\omega_{\mathrm{ir}}t)+1$, which is measurement specific. However, for an Hahn echo experiment, the filter function converges independent of measurement details and $\omega_\mathrm{ir}$ is effectively set by $1/T_{2E}$, with $\xi_{\mathrm{E}} = \mathrm{ln}2$.

{\bf Extracting the flux noise amplitude: } We perform measurements of $\TE$ over the entire range of flux-biases in order to extract the flux noise amplitude for each qubit. Away from the zero- and half-flux sweet spots, the qubit is sensitive to first-order flux noise as given by 
\begin{equation}
     \GammaphiE^{\Phi}  = \left| \frac{\partial\omega_{01}}{\partial\Phiext} \right| A_{\Phi} \sqrt{\mathrm{ln} 2}.
     \label{eq:Gamma_phiE_fluxnoise}
\end{equation} 
In Fig. \ref{fig:S2}, we plot the measured echo dephasing rate $\GammaphiE$ as a function of qubit frequency, and the expected first-order flux noise dephasing rate (solid line), given a particular $A_{\Phi}$ and the qubit-specific $\partial\omega_{01}/\partial\Phiext$. The extracted flux noise amplitudes for the six qubits we measured fall in a range of $A_\Phi \approx 3-5 \ \mu\Phi_0 / \sqrt{\mathrm{Hz}}$ (see Table \ref{tab:Fluxonium Parameters}), in agreement with values reported in other experiments \cite{braumuller_characterizing_2020,yan_flux_2016,sun_characterization_2023-1}. 

{\bf Flux noise contribution to Ramsey dephasing: }
We use a minimal model to fit the Ramsey dephasing rate shown in \figlbl\figref{fig:3}{b}, which includes contributions from CQPS and flux noise, $1/\TphiR = \GammaphiR = \sqrt{(\GammaCQPS)^2 + (\GammaphiR^{\Phi})^2}$. The CQPS dephasing is given by Eqn. \ref{eq:GammaCQPS}. The contribution from flux noise is has the functional form
\begin{equation}
    \GammaphiR = A_{R1} \left| \frac{\partial\omega_{01}}{\partial\Phiext} \right| 
\end{equation}
where $A_{R1} = A_{\Phi}\sqrt{\xi_{\mathrm{R}}} \approx A_{\Phi } \sqrt{\mathrm{ln}(1/\omega_{\mathrm{ir}}t)+1}$. Although the $\TE$ data provides an independent measure of $A_{\Phi}$, the IR cutoff was not held constant for all Ramsey measurements, which results in variable $A_{R1}$. Instead of allowing $A_{R1}$ to be a free parameter for each fit, we choose a proportionality constant $A_{R1} = 4 A_{\Phi} \sqrt{\mathrm{ln}2}$.  
In the case where $\GammaphiR$ is dominated by first-order flux noise, the ratio of $\GammaphiR/\GammaphiE \approx 4$ is determined by the duration of the Ramsey experiment and is similar to values observed in other works \cite{bylander_noise_2011}.

\section{Noise power spectral density}
\label{app:noisePSD}
{\bf Repeated Ramsey measurements: }
To extract the PSD of qubit frequency fluctuations, we perform a series of repeated Ramsey measurements. Each Ramsey sequence consists of 100 equally-spaced delay times between the two $\pi/2$ pulses, 500 averages for each delay, and a trigger period of 1 ms between successive measurements, for a total measurement time of 58 seconds per point with data transfer overhead time included.  This protocol is repeated 1000 times to give $\sim 3$ decades of sampled frequencies.

{\bf Q6 flux noise amplitude from noise PSD: }
In \figlbl\figref{fig:4}{b}, we show that the frequency fluctuations for Q6 as extracted from repeated Ramsey measurements follow a noise power spectral density of the form $S_f(\omega) = 2\pi M/|\omega|$. The qubit frequency fluctuations can be converted into a measure of flux noise by taking into account the qubit sensitivity to external flux, 
\begin{equation}
    S_f(\omega) = S_\Phi(\omega) \left( \frac{df_{01}}{d\Phiext}\right)^2 
\end{equation}
with a flux noise amplitude given by $A_\Phi = \sqrt{M/(df_{01}/d\Phiext)^2}$. The repeated Ramsey measurements for Q6 in \figlbl\figref{fig:4}{b} were performed at $\Phiext = 0.42 \Phi_0$. From the fit of $S_f(\omega)$, we obtain $A_\Phi = 5.9 \ \mu\Phi_0  / \sqrt{\mathrm{Hz}}$, similar to the value for this qubit from $\TE$  measurements.

\section{Modeling parity switching in the JJA}
\label{app:qp_simulations}
For the noise simulations in \figlbl\figref{fig:5}{} of the main text, we consider 10 quasiparticles (charge $1e$), with randomly initialized locations in the array at $t=0$. 
This choice of quasiparticle population in the array corresponds to a quasiparticle fraction $x_\mathrm{QP}=4\times10^{-7}$,
based on an estimate of the total superconductor volume used in the JJ array of $\approx 67\mu \mathrm{m}^2\times0.1\ \mu \mathrm{m}=6.7\ \mu\mathrm{m}^3$,
and a Cooper-pair density of $n_\mathrm{CP}\approx4\times10^6~\mu \mathrm{m}^{-3}$ \cite{wang_measurement_2014}. 
Within the simulation, we allow quasiparticle tunneling with equal probability to the left or right, such that if the charge parity changes on the $j$-th island, a corresponding change of opposite sign is implemented on the adjacent $j-1$ or $j+1$ island. We enforce periodic boundary conditions, in order to keep the number of quasiparticles in the array fixed. To cover a similar range of frequencies probed in the repeated Ramsey measurements, we calculate the number of quasiparticles on each island, $\nQP{}_{,j}(t)$, in 500 $\mu$s steps for $10^4$ seconds. Although, in principle $\delta n_{g,j}$ can change in time, we find that a fixed offset is sufficient to match the data measured from Q1 over the frequency range of interest. The value of $\tauQP =\ 10$ ms used in the simulations for \figlbl\figref{fig:5}{} of the main text is meant to be illustrative. Moreover, to avoid numerical artifacts arising from a particular configuration of initial quasiparticle positions in the array, we repeat the entire simulation process 10 times and take the average of the noise PSD to generate the trace in \figlbl\figref{fig:5}{}. From the repeated Ramsey measurements, $S_f\left(\omega\right)$ is found to be frequency independent below 10$^{-2}$~Hz. Additionally, from the increase of $\TphiE$ compared to $\TphiR$, one can infer that the PSD is reduced near $1/\TphiE\sim\ 10^4$ Hz.  The choice of $\tauQP =\ 10$ ms places the roll-off frequency for the noise specturm between these reference points and is consistent with the measured data.  These simulations leave open the possibility of future studies to probe the noise spectrum at higher frequencies, as well as investigate intentional changes to the array junctions to modify $\tauQP$.  

\section{Structure factor}
\label{app:structure factor}
\begin{figure}
    \centering
    \includegraphics{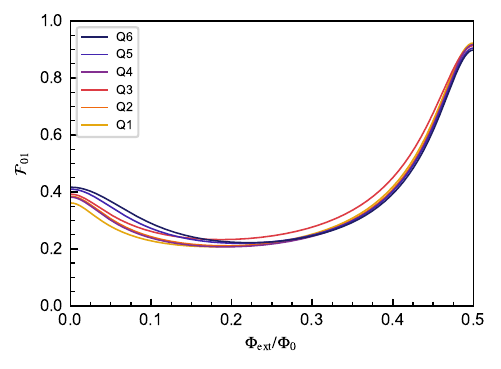}
    \caption{Structure factor versus flux for all qubits using values reported in Table \ref{tab:Fluxonium Parameters}.}
    \label{fig:S4}
\end{figure}

We plot the flux dependence of the structure factor $\mathcal{F}_{01}$ (\eqlbl~\ref{eq:structurefactor}) for all qubits in \figlbl\figref{fig:S4}{}. A larger $\mathcal{F}_{01}$ value makes the qubit more susceptible to CQPS dephasing, as seen here around half-flux.

\section{Fluxon-transition broadening due to coherent quantum phase slips}
\label{app:theory}

\subsection{Phase-slip representation and perturbation theory}
\label{subsub:perturbation theory}
In this section, we present a self-contained theory to derive the broadening of the fluxon transition due to coherent quantum phase slips. More precisely, we use Bloch theory to arrive at the fluxonium Hamiltonian in a representation that reveals the phase-slip rate due to the high impedance small junction first, and propose an extension of this model to include phase slips in the junction array. We then use the full-circuit Hamiltonian to calculate a correction to the fluxon-transition frequency due to the combined effect of charge fluctuations and a ``slippery" superinductance, which can then be used to compute an effective pure-dephasing rate associated with this noise channel.

We begin with the fluxonium Hamiltonian [\cref{eq:fluxoniumHamiltonian}], which we rewrite here for simplicity
\begin{equation}
    \Hop = 4 \EC (\nop-\ngate)^2 -\EJ\cos\phiop + \frac{\EL}{2}(\phiop+\phiext)^2.
    \label{eq:H_0}
\end{equation} 
Note that we account for the offset-charge bias~$\ngate$, which is typically excluded in the literature. This omission is in most cases justified, as a time-independent~$\ngate$ can be removed from the Hamiltonian by a gauge transformation~\cite{koch_charging_2009}. However, we show below that preserving the parametric $\ngate$-dependence of~\cref{eq:H_0} throughout a few theoretical steps leads to a form of the fluxonium Hamiltonian that highlights the impact of charge noise, which we focus on in this work.

A Bloch basis has been used before to study the~$\EL\to 0$ limit of the fluxonium circuit~\cite{koch_charging_2009}. Our derivation largely follows the pioneering work of \textit{Koch et al.}, but it differs from it in the treatment of~$\ngate$. It will become clear below that this seemingly small change is crucial for the purposes of our work. 

We split the fluxonium Hamiltonian into two terms, with one of them being the $2\pi$-periodic portion
\begin{equation}
    \Hopp = 4 \EC (\nop-\ngate)^2 -\EJ\cos\phiop.
    \label{eq:H_p}
\end{equation}
While this Hamiltonian looks identical to that of charge (or transmon) qubit, no $2\pi$-periodic boundary conditions are imposed to the eigenstates of~\cref{eq:H_p}. This is because the fluxonium degree of freedom~$\phiop\in(-\infty,\infty)$ is noncompact and the cosine-potential minima located at~$2\pi l$ with~$l\in\mathds{Z}$ are distinguishable. According to Bloch theorem, $\Hopp$ eigenstates are quasi-periodic and have the general form
\begin{equation}
    \psi_b^k(\phi)=e^{ik\phi}u_b^{k}(\phi),
    \label{eq:H_p eigenstates}
\end{equation}
where, $u_b^{k}(\phi)$ are~$2\pi$-periodic function of the phase, $k\in[-1/2,1/2)$ is a continuous quantum number analogous to the ``crystal momentum" in solid state systems, and~$b$ is an integer-valued quantum number which represents the band index. The Bloch eigenstates satisfy the orthogonality relation~$\langle\psi_b^k|\psi_{b'}^{k'}\rangle=\delta_{b b'}\delta(k-k')$, with the first $\delta$ being of the Kronecker variety and the second denoting a Dirac delta function in the $k$ basis. 

The $2\pi$-periodic part of the eigenstates in~\cref{eq:H_p eigenstates} is determined by the eigenvalue equation~$\Hopp\psinkket=\hbar\omega_b^k\psinkket$, which can be rewritten as
\begin{equation}
    \left[4 \EC (\nop + k - \ngate)^2 -\EJ\cos\phiop\right]|u_b^k\rangle=\hbar\omega_b^k|u_b^k\rangle,
    \label{eq:unk eigenvalue problem}
\end{equation}
where~$u_b^{k}(\phi)=\langle \phi|u_b^{k}\rangle$. Note that~\cref{eq:unk eigenvalue problem} is subject to~$2\pi$-periodic boundary conditions~$u_b^{k}(\phi)=u_b^{k}(\phi+2\pi)$; in other words, $|u_b^{k}\rangle$ are the eigenstates of a charge qubit with~$\EC$ and~$\EJ$ denoting the single-electron charging and Josephson energies, respectively, and effective offset-charge bias~$\ngate-k$. 

In this new basis, \cref{eq:H_p} takes the diagonal form:
\begin{equation}
    \Hopp = \sum_{b\geq 0}  \int_{-\frac{1}{2}}^{\frac{1}{2}}  \hbar\omega_b^k |\psi_{b}^k\rangle \langle \psi_{b}^k| dk.
    \label{eq:H_p diagonal}
\end{equation}
For any nonzero value of~$\EL$, the quadratic potential associated with the inductance breaks the~$2\pi$-periodicity of the Josephson potential. As a consequence, the Bloch-basis indices~$k$ and~$b$ are no longer good quantum numbers. Regardless, the Bloch basis remains useful to analyze the fluxonium Hamiltonian, provided the representation for the phase operator~\cite{koch_charging_2009}
\begin{equation}
    \phiop \to i\partial_k + \hat{\Omega}.
    \label{eq:phi operator}
\end{equation}
The first term in~\cref{eq:phi operator}, $i\partial_k$, couples Bloch eigenstates with different crystal momenta, generating a translation in~$k$-space within a single Bloch band. In addition, $\hat{\Omega}$ is a $k$-independent interband potential that couples Bloch states belonging to different bands~($b\neq b'$). With these definitions the fluxonium Hamiltonian takes the form
\begin{equation}
    \Hop = \sum_{b\geq 0}  \int_{-\frac{1}{2}}^{\frac{1}{2}}  \hbar\omega_b^k |\psi_{b}^k\rangle \langle \psi_{b}^k| dk + \frac{\EL}{2}\left(i\partial_k + \hat{\Omega} +\phiext \right)^2,
    \label{eq:fluxonium Hamiltonian Bloch representation}
\end{equation}
where~$k$ ($b$) must now be regarded as a compact (integer-valued) degree of freedom~$k\to \hat{k}$ ($b\to \hat{b}$). Note that~$\hat b$ is analogous to a photon number--an interpretation that will become clearer below.

To further simplify our analysis, we rewrite~\cref{eq:fluxonium Hamiltonian Bloch representation} in the basis where the conjugate-charge~$\mop$ associated with~$\hat k$ is a diagonal operator, i.e., $\mop = \sum_{m\in\mathds{Z}} m|m\rangle\langle m|$. This change of basis is implemented by a Fourier unitary~$\hat U_\mathrm{ps}$, under which~\cref{eq:H_p diagonal} transforms as
\begin{equation}
    \Hopp \to \sum_{b\geq 0} \left[\hbar\bar{\omega}_b + \sum_{l> 0} \frac{\epsilon_\mathrm{ps}(b,l)}{2} e^{i2\pi (\hat k-\ngate)l} + \mathrm{H.c.}\right]|b\rangle\langle b|.
    \label{eq:H_p phase slip}
\end{equation}
Here, $\hbar\bar{\omega}_b$ is the average energy of the~$b$th Bloch band, $\epsilon_\mathrm{ps}(b,l)$ is the $l$th phase-slip energy associated with band~$b$, and~$|b\rangle\langle b|$ is a band projector. [The meaning of the name given to~$\epsilon_\mathrm{ps}(b,l)$ and its functional form in terms of the fluxonium parameters are discussed below.] To write down the general Fourier decomposition in~\cref{eq:H_p phase slip}, we used the fact that the eigenvalues~$\hbar\omega_b^k$ are 1-periodic functions of~$k-\ngate$. [Such a periodicity is a well-known property of the charge-qubit eigenenergies, and can be shown in two steps: \textit{i}) implement the translation~$k-\ngate \to k - \ngate \pm 1$ in the eigenvalue problem~\cref{eq:unk eigenvalue problem}, and \textit{ii}) displace the charge operator as~$\nop \to \nop \mp 1$, returning the eigenvalue equation to its original form.] The phase operator in~\cref{eq:phi operator} transform as
\begin{equation}
    \phiop \to -2\pi \hat m + \hat{\Omega}_\mathrm{ps},
    \label{eq:phi operator phase slip}
\end{equation}
with~$\hat{\Omega}_\mathrm{ps} = \hat{U}^\dagger_\mathrm{ps} \hat{\Omega}\hat{U}_\mathrm{ps}$ leading to
\begin{equation}
    \begin{split}
        \Hop &= \frac{\EL}{2}\left(\phiext -2\pi \hat m + \hat{\Omega}_\mathrm{ps}\right)^2 \\
&+ \sum_{b\geq 0}\sum_{l>0} \frac{\epsilon_\mathrm{ps}(b,l)}{2} e^{-i2\pi\ngate l} (\mop^+)^l |b\rangle\langle b| + \mathrm{h.c.},
    \end{split}
    \label{eq:phase slip Hamiltonian}
\end{equation}
where we introduced the phase-slip operators~$\mop^{\pm}=e^{\pm i 2\pi \hat{k}}$, with~$\mop^{\pm}|m\rangle = |m\pm 1\rangle$, and omitted a constant energy term. The qualifier ``phase-slip" is motivated by~\cref{eq:phi operator phase slip}: a displacement~$m\to m + l$ with~$l\in \mathds{Z}$ leads to a shift of the fluxonium phase by~$2\pi l$ units (disregarding the effect of the interband potential). The Hamiltonian in~\cref{eq:phase slip Hamiltonian} has a simple interpretation: in fluxonium, the persistent current or ``fluxon" states~$|m\rangle$, which are eigenstates of the inductive or ``loop" Hamiltonian, are coupled by the phase-slip potential that describes the effect of the Josephson junction. In other words, the Josephson junction behaves as a tunneling element for fluxons in and out of the fluxonium loop. 

We use this qualitative picture to propose a generalization of~\cref{eq:phase slip Hamiltonian} that can account for the internal degrees of freedom of the superinductance implemented by a junction array. In this scenario, the $j$th Josephson junction of the array, with~$j\in[1,N]$, contributes with (\textit{i}) a $b_j$-band fluxon-tunneling amplitude~$\epsilon_{\mathrm{ps},j}(b_j,l)e^{-i2\pi\eta_{g,j} l}/2$, which depends on the array-junction parameters~$E_{\mathrm{C},j}$ and~$E_{\mathrm{J},j}$ and 
an associated charge parameter ~$\eta_{g,j}$, and (\textit{ii}) an independent contribution to the interband potential~$\hat\Omega_{\mathrm{ps},j}$. Here, ~$j=0$ corresponds to the fluxonium high-impedance JJ with parameters~$\EC$ and~$\EJ$, and an offset charge across the junction $n_g = n_{g,0} = \eta_{g,0}$, referenced to ground as depicted in \figlbl\figref{fig:1}{b}. The physical offset charges associated with each superconducting island in the circuit factor into the accumulated phase of the fluxon tunneling amplitude. The effective charge parameters that enter into the Hamiltonian depend on the (inverted) full capacitance matrix, thereby making $\eta_{g,j}$  nonlocal with respect to $n_{g,j}$ (see \cite{di_paolo_efficient_2021} for details).
The generalization of the Bloch-basis treatment for fluxonium that we present here ultimately leads to a Hamiltonian that is consistent with defining $\eta_{g,j} = \sum_{k=0}^{j-1} n_{g,k}$.
In the main text, we motivate how this geometric phase arises from the Aharonov-Casher effect, and depends on the cumulative charge between the small junction and the $j$-th junction in the array.

This analysis leads us to the effective microscopic fluxonium Hamiltonian 
\begin{equation}
    \begin{split}
        \Hop_\mathrm{micro} &\approx\frac{\EL}{2}\left(\phiext -2\pi \hat m + \sum_{j\geq 0}^{N}\hat{\Omega}_{\mathrm{ps},j}\right)^2 \\
        &+ \sum_{j=0}^N\sum_{b_j\geq 0}\sum_{l>0} \frac{\epsilon_\mathrm{ps}(b_j,l)}{2} e^{-i2\pi \eta_{g,j} l} (\mop^+)^l |b_j\rangle\langle b_j| \\
        &+ \mathrm{H.c.},
    \end{split}
    \label{eq:phase slip Hamiltonian micro}
\end{equation}
where $|b_j\rangle\langle b_j|$ is a diagonal band operator involving the new band-occupation index~$b_j$. 
This microscopic model assumes that the effective inductive energy that each junction in fluxonium is subject to--due to the combined effects of all other junctions in the circuit--is the same for all junctions. This approximation is justified in the fluxonium limit, where the number of junctions is large and the contribution of an individual junction to the array inductance is small compared to the total value. 

\Cref{eq:phase slip Hamiltonian micro} largely increases the dimensionality of our fluxonium model to a point where further analytical treatment would be impractical. Thus, to make analytical progress we introduce two more approximations. First, we assume that the interband potential associated with the~$j$th array junction does not contribute to the low-energy physics of fluxonium. This approximation is justified by noticing that~\cite{koch_charging_2009}
\begin{equation}
    \langle b_j|\hat\Omega_{\mathrm{ps},j}|0_j\rangle\propto \frac{1}{\omega^{k}_{0_j}-\omega^{k}_{b_j}},
\end{equation}
where~$|0_j\rangle\langle b_j|$ is an interband jump operator, and~$\hbar\omega^{k}_{b_j}$ are the eigenvalues in~\cref{eq:unk eigenvalue problem} but associated with the~$j$th array junction, i.e., $(\EC,\EJ,\ngate)\to (E_{\mathrm{C},j},E_{\mathrm{J},j},\eta_{g,j})$. Assuming that the plasma frequency~$\textstyle \sqrt{8 E_{\mathrm{C},j}E_{\mathrm{J},j}}$ of the array junctions is large, the interband potential associated with the array junctions contributes negligibly to the low-energy physics of the qubit. In other words, this approximation amounts to a partial-trace operation that sets~$|b_j\rangle\langle b_j|\to 0$ for~$j\geq 1$, and~$|b_0\rangle\langle b_0|\to 1$, recovering a low-dimensional fluxonium Hamiltonian that involves the original degrees of freedom~$(\hat m,\hat b)$. The second approximation concerns the ratio~$E_{\mathrm{J},j}/E_{\mathrm{C},j}$, which determines the zero-point phase fluctuations across the array junctions, that we assume to be small compared to~$2\pi$ (or~$E_{\mathrm{J},j}/E_{\mathrm{C},j}\gg 1$).  We show in~\cref{subsub:phase slip energies} that this also implies
\begin{equation}
    |\epsilon_{\mathrm{ps},j}(b_j,1)|\gg|\epsilon_{\mathrm{ps},j}(b_j,l>1)|,
    \label{eq:phase slip energy relation}
\end{equation}
or~$|\epsilon_{\mathrm{ps},j}(b_{j\geq 1},l>1)|\approx 0$ for~$j\geq 1$. With these approximations, our fluxonium model reduces to
\begin{equation}
    \begin{split}
        \Hop_\mathrm{micro} &\approx \frac{\EL}{2}\left(\phiext -2\pi \hat m +\hat{\Omega}\right)^2 \\
        &+ \sum_{b\geq 0}\sum_{l>0} \frac{\epsilon_\mathrm{ps}(b,l)}{2} e^{-i2\pi n_{g,0} l} (\mop^+)^l |b\rangle\langle b| + \mathrm{H.c.} \\
        &+ \sum_{j\geq 1}^N \frac{\epsilon_{\mathrm{ps},j}}{2} e^{-i2\pi \eta_{g,j}} \mop^+ + \mathrm{H.c.},
    \end{split}
    \label{eq:phase slip Hamiltonian micro simp}
\end{equation}
where we simplified the notation as~$\epsilon_\mathrm{ps}(0_j,1)\to \epsilon_{\mathrm{ps},j}$, and set~$|0_j\rangle\langle 0_j |\to 1$ to simplify notation. 

The first two lines in~\cref{eq:phase slip Hamiltonian micro simp} correspond to the fluxonium model in~\cref{eq:phase slip Hamiltonian}, and can be put in diagonal form in terms of its eigenstates~$|\psi_\alpha\rangle$ and respective eigenenergies~$\epsilon_\alpha$. In practice, the offset charges~$n_{g,j}$ are subject to low-frequency fluctuations, broadening the qubit energy transition and leading to an effective pure-dephasing rate. To calculate such a rate, we require an estimation of the energy shifts~$\delta \epsilon_\alpha$ for $\alpha\in[0,1]$ due to the perturbation (third line) in~\cref{eq:phase slip Hamiltonian micro simp}. To first order in the phase-slip amplitudes~$\epsilon_{\mathrm{ps},j}$, the corrections to the eigenenergies are
\begin{equation}
        \delta \epsilon_\alpha =\sum_{j\geq 1}^N \frac{\epsilon_{\mathrm{ps},j}}{2} e^{-i2\pi\eta_{g,j}} \langle \psi_\alpha| \mop^+|\psi_\alpha\rangle + \mathrm{c.c.}
    \label{eq:perturbation theory correction_app}
\end{equation}
The matrix elements~$\langle \psi_\alpha| \mop^+|\psi_\alpha\rangle$ can be computed in an exact way by diagonalizing the fluxonium Hamiltonian in fluxon basis [\cref{eq:phase slip Hamiltonian}], where the operators~$\mop^{\pm}$ are well defined. However, numerically defining the fluxonium Hamiltonian in this basis is not as straightforward as in other bases (such as phase or Fock basis) and not a choice of preference in the literature. Thus, approximating~\cref{eq:perturbation theory correction} using an alternative description of fluxonium is useful. Neglecting the interband potential, we can write
\begin{equation}
    \langle \psi_\alpha| \mop^+|\psi_\alpha\rangle \approx  \langle \psi_\alpha | e^{-i2\pi\nop} |\psi_\alpha\rangle,
\end{equation}
which links the matrix element of the phase-slip operator with that of a displacement operator in phase space. Subtracting the ground- and first-excited state corrections in~\cref{eq:perturbation theory correction_app}, we arrive to~\cref{eq: fab} in the main text, previously stated in~\cite{manucharyan_evidence_2012}. This completes our derivation of the perturbative effects of coherent quantum phase slips on the fluxon (qubit) transition of fluxonium. 

\subsection{Single and correlated phase-slip amplitudes}
\label{subsub:phase slip energies}

\begin{figure}[ht!]
\centering
\includegraphics[width=0.475\textwidth]{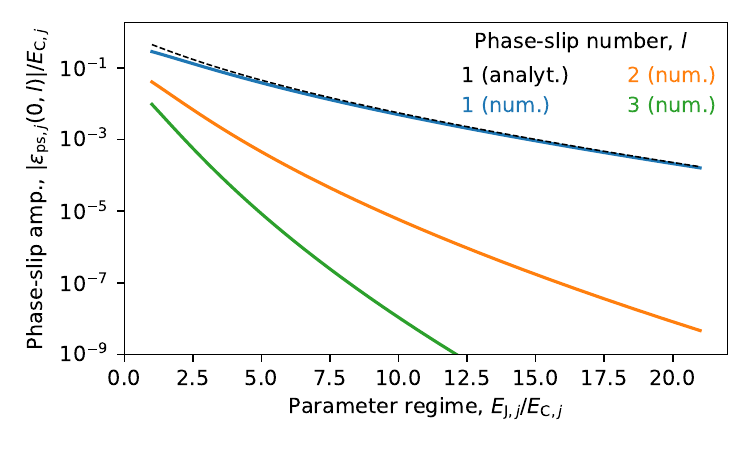}
\caption{\textbf{Junction-array phase-slip energies.} Phase-slip amplitudes (in units of the charging energy~$E_{\mathrm{C},j}$) extracted from a Fourier-series decomposition of the ground-state energy, $\hbar\omega_{0_j}^k$, as a function of the ratio~$E_{\mathrm{J},j}/E_{\mathrm{C},j}$. The dashed line shows the analytical value predicted by~\cref{eq:transmon ground-state energy coefficient}. Note that $\epsilon_{\mathrm{ps},j}(b_j,1)\gg \epsilon_{\mathrm{ps},j}(b_j,l>1)$ for typical~$E_{\mathrm{J},j}/E_{\mathrm{C},j}$ ratios, justifying the single phase-slip approximation for the array junctions.}
\label{fig:phase_slip_energy}
\end{figure}

The phase-slip energies are determined by Fourier-transforming the eigenvalues~\cref{eq:unk eigenvalue problem} (solved for the array junctions) as a function of the ``bias"~$k$. In the transmon regime ($E_{\mathrm{J},j}\gg E_{\mathrm{C},j}$), it is well-known that the ground-state energy can be approximated by~\cite{koch_charge-insensitive_2007}
\begin{equation}
    \hbar\omega_{0_j}^k \approx - \epsilon_{0,j} \cos[2\pi (k-\eta_{g,j})],
\end{equation}
where
\begin{equation}
    \epsilon_{0,j} = 4\sqrt{2/\pi}\sqrt{8 E_{\mathrm{C},j}E_{\mathrm{J},j}}(2 E_{\mathrm{C},j}/E_{\mathrm{J},j})^{-1/4}\exp^{-\sqrt{\frac{8E_{\mathrm{J},j}}{E_{\mathrm{C},j}}}}.
    \label{eq:transmon ground-state energy coefficient}
\end{equation}
This analytical estimate is shown in~\cref{fig:phase_slip_energy} alongside higher-order Fourier components for the~$b_j=0$ band potential. \Cref{eq:transmon ground-state energy coefficient} is a very good approximation for most junction parameters, improving in the transmon regime. We also note that the amplitudes of correlated phase slips $l>1$ are much smaller than that of single phase slips, justifying the approximation introduced in~\cref{eq:phase slip energy relation}.

\twocolumngrid

\clearpage
\newpage 

\bibliography{cqps_arxiv_final.bib}
\end{document}